\documentclass[letterpaper]{article}

\usepackage{fullpage,authblk}
\usepackage{amssymb,amsfonts,amsmath,enumerate}
\usepackage[bookmarks, colorlinks=true, plainpages = false, citecolor = green, urlcolor = blue, filecolor = blue]{hyperref}
\usepackage{graphicx}
\usepackage{multirow}
\usepackage{appendix}

\newcommand{\bA}{{\mathbf A}}
\newcommand{\bv}{{\mathbf v}}
\newcommand{\ba}{{\mathbf a}}
\newcommand{\bc}{{\mathbf c}}
\newcommand{\br}{{\mathbf r}}
\newcommand{\bx}{{\mathbf x}}
\newcommand{\by}{{\mathbf y}}

\newcommand{\bX}{{\mathbf X}}

\newcommand{\bY}{{\mathbf Y}}
\newcommand{\bu}{{\mathbf u}}
\newcommand{\bW}{{\mathbf W}}
\newcommand{\bV}{{\mathbf V}}
\newcommand{\bR}{{\mathbf R}}

\newcommand{\be}{{\mathbf e}}
\newcommand{\bw}{{\mathbf w}}
\newcommand{\bb}{{\mathbf b}}

\newtheorem{theorem}{Theorem}

\newtheorem{remark}{Remark}

\DeclareMathOperator*{\argmin}{arg\,min}

\begin{document}

\title{
On Explaining the Surprising Success of Reservoir Computing  Forecaster of Chaos? \\ The Universal Machine Learning  Dynamical System with Contrasts to VAR and DMD
}


\def\correspondingauthor{\footnote{Corresponding author: bolltem@clarkson.edu}}
\author[1,2]{Erik Bollt \correspondingauthor{}}

\affil[1]{Department of Electrical and Computer Engineering, Clarkson University, Potsdam, NY 13699, USA}
\affil[2]{Clarkson Center for Complex Systems Science ($C^3S^2$), Potsdam, NY 13699, USA}

\date{}
\maketitle
\begin{abstract}
{\color{black} Machine learning has become a widely popular and successful paradigm, including in data-driven science and engineering.  A major application problem is data-driven forecasting of future states from a complex dynamical.  Artificial neural networks (ANN) have evolved as a clear leader amongst many machine learning approaches, and recurrent neural networks (RNN) are considered to be especially well suited for forecasting dynamical systems.  In this setting, the echo state networks (ESN) or reservoir computer (RC) have emerged for their simplicity and computational complexity advantages.  
Instead of a fully trained network, an RC trains only read-out weights by a simple, efficient least squares method.  What is perhaps quite surprising is that nonetheless an RC succeeds  to make high quality forecasts, competitively with more intensively trained methods, even if not the leader.   There remains an unanswered question as to why and how an RC works at all, despite randomly selected weights.  To this end, this work analyzes a further simplified RC, where the internal activation function is an identity function.  Our simplification is not presented for sake of tuning or improving an RC, but rather for sake of analysis of what we take to be the surprise being not that it doesn't work better, but that such random methods work at all.   We explicitly connect the RC with linear activation and linear read-out to well developed time-series literature on vector autoregressive averages (VAR) that includes theorems on representability through the WOLD theorem, which already perform reasonably for short term forecasts.  In the case of a linear activation and now popular quadratic read-out RC, we explicitly connect to a nonlinear VAR (NVAR), which performs quite well.  Further, we associate this paradigm to the now widely popular dynamic mode decomposition (DMD), and thus these three are in a sense different faces of the same thing.  We illustrate our observations in terms  of  popular benchmark examples including Mackey-Glass differential delay equations and the Lorenz63 system.}
\end{abstract}

\noindent Key Words: linear reservoir computing, RC, neural network, recurrent neural network, RNN, machine learning, vector autoregression, VAR, Wold theorem, dynamic mode decomposition, DMD. 

\medskip

\noindent {\bf 
The power and success of artificial neural networks has been profound across many disciplines, including in dynamical systems. A leader amongst methodologies for forecasting has been the recurrent neural network (RNN) for aspects of memory.  However, because of the large number of parameters to train to data observations, and likewise the nonlinear nature of the associated optimization process, the training phase can be computationally extremely intensive.    The echo-state, reservoir computing  (RC) concept is a significant simplification where only the output weights are trained and in a manner that allows for a straight forward and cheap least squares method.  The rest of the weights, those of the input layer and those of inner layers are simply selected randomly.  It is clear that this would be cheaper to train, but what is not clear and perhaps a surprise is that it would work at all, but work it does.  With a simplification of the concept to allow for a linear activation function, while the performance is not quite as good it does still work, and now we are able to analyze in detail the role of the randomly selected parameters and how there is still freedom in fitting a well defined time-series forecasting model, which in fact is equivalent to the well developed theory of vector autoregression (VAR).  Within the VAR and related VMA theory we recall the Wold theorem that allows us to discuss representation, and now as we show it is relevant to the RC for machine learning.  Also, with this description, we are able to connect to the recently highly popular DMD concept.  {\color{black} While we do observe that the fully linear version of the RC, and so corresponding VAR, does make reasonable short term forecasts, a linear RC with quadratic readout significantly improves forecasts and even apparently once errors do occur, they seem more true to the true nature of the original system.  In the spirit of the linear RC plus linear readout yields a VAR, we show that linear RC with a (Hadamard) quadratic readout  yields a nonlinear VAR (NVAR) that includes monomials of all quadratic forms.}}

\section{Introduction}

Artificial neural networks (ANN) have emerged as a core and powerful technology  in machine learning \cite{de1996neural,levine2018introduction,  marsland2015machine, michie1994machine,  nelson1991practical} that is well suited for the supervised learning in data-driven science and engineering, specifically including for forecasting problems in complex dynamical systems \cite{funahashi1993approximation, kumpati1990identification,huang2003neural,li2017extended,brunton2019data,langkvist2014review,hartman2017deep}.
However, the most straight  forward feedforward ANN with back propagation for training concepts can be extremely expensive to optimize to the data, even considering important recent innovations such as stochastic gradient descent or hardware break throughs such as GPU-based  processing.  Recurrent neural network concepts, (RNN) are especially suitable for temporal data from a dynamical system \cite{funahashi1993approximation,kimura1998learning,bailer1998recurrent,barbounis2006long,choi2017using,rather2015recurrent}, as they naturally embed temporal information, and especially the long short term memory (LSTM) approach demonstrate excellent fidelity, \cite{vlachas2019forecasting,hochreiter1997long,vlachas2018data,choi2017using,chattopadhyay2019data,yeo2019deep}, but these are especially expensive to fully train, \cite{pascanu2013difficulty}.

The reservoir computing (RC) \cite{jaeger2004harnessing,  lukovsevivcius2009reservoir, verstraeten2007experimental} and the closely related echo state network (ESN) \cite{jaeger2001echo, lukovsevivcius2012practical} and liquid state machine (LSM) \cite{maass2002real, grzyb2009model} have emerged as a special variant of an RNN,  
where only the output layer is trained rather than the entire network of weights.  As such, this requires only a simple and efficient least squares estimation, rather than the more expensive full nonlinear optimization associated with a fully training an RNN.  Nonetheless, and perhaps a most surprising outcome is that despite this gross simplification, the forecasting capability can still be competitive even for chaotic or spatiotemporally complex problems \cite{vlachas2019forecasting,pathak2018model, zimmermann2018observing,lu2018attractor,canaday2018rapid,carroll2019network,gauthier2018reservoir}.  Specifically, an RC thrives when a full state observation is available, while fuller and more expensive variants of RNN,  especially the  LSTM would considered higher perming, especially when only a reduced variable set is available \cite{vlachas2018data,vlachas2019forecasting}.  Still, the RC are popular, surely because of their simplicity to train, and perhaps in part because of their undeniable even if surprising fidelity.

{\color{black} The purpose of this work is to offer at least a partial explanation as to how an RC can be such a successful and general universal dynamical system for forecasting such a wide array of systems despite randomly ``trained" read-in and inner weights.  In other words, we work to better understand,  "where does the randomness go?"  The purpose of this paper is not specifically to build a new method, or to improve the current method, but to explain what is perhaps a surprising that the RC method works at all. In so doing, we challenge the concept with a simplified linear activation function version for sake that this allows our simplified analysis throughout, and even if this version has reduced fidelity, we show it does still have theoretic reasons it still works which we are now in a position to describe in detail. Our simplification to a linear activation function also allows us to explicitly write the RC as a VAR,  serving as a bridge to the more theoretically well established time series theory, and also to DMD.  Nonetheless, we do describe a simple scenario where the linear RC is expected to perform no worse than the widely used nonlinear counterpart in terms of efficiency, but the linear version we can now show in detail without ambiguity when  there will be a good forecasting version by tying it to the representation theory of VAR as seen through the Wold theorem.
While the linear activation reservoir and linear read-out is explicitly connected to a VAR and results at least for short term forecasts are reasonable, we also show that linear reservoir with quadratic read-out (as quadratic read-out has become popular \cite{pathak2018model}) is equivalent to a NVAR and this turns to perform quite well.
}

There have been few explanations as to how despite the random construction, an RC works so well, but notably \cite{gonon2019reservoir, buehner2006tighter, darmon2019information}.  Usually instead we find in the literature a collection of descriptions as to how to choose random networks as the inner layers, regarding sparsity \cite{song2010effects, griffith2019forecasting}, or regarding design of the  spectral radius for linear stability \cite{carroll2019network, jiang2019model,griffith2019forecasting} and the echo property, \cite{gallicchio2018chasing}.   {\color{black} An especially strong result comes from study of Volterra series, by Boyd and Chua \cite{boyd1985fading} where it was proven that a finite-dimensional linear dynamical with a nonlinear read-out, even a polynomial read-out, can approximate very general signals.   A new analysis by Hart, Hook and Dawes show that regularized echo state machines make good universal approximators of ergodic dynamical systems \cite{hart2020echo} and furthermore give generalized embedding results \cite{hart2020embedding} reminiscent of the classical Taken's embedding theorem \cite{takens1981detecting}.    Also recently it has been shown that fading memory leads to universality in Gonan and Ortega, \cite{gonon2020fading}. }

In the spirit of still incomplete theoretical basis as to the underlying success of RC, 
 we allow a simplified version of RC with linear activation functions for which we are able to more fully identify the inner workings of how the RC can be a universal forecasting machine, even for  time-series from complex and chaotic dynamical systems.  We show that by this simplification the RC still works, albeit with reduced performance quality, but nonetheless the purpose here being theoretical explanation of how such a simple system of only training the read-out is possible.  We offer this variant as a theoretical construction.  By this interpretation,  we will also be able to connect the RC concept to other theoretically more matured theories.  Specifically, the theory of autoregression (AR) from time-series analysis and moving averages (MA), and together called ARMA \cite{pandit1983time,chen1995analysis,box1970time,pena2011course,tiao1983consistency,serio1994autoregressive} , are founded on the Wold theorem \cite{wold1954study} that we show  are directly related to the RC concept.  The vector formulation of these \cite{qin2011rise,lutkepohl2005new}, called vector autogregression (VAR) and vector moving averages (VMA) are also connected by a corresponding Wold theorem. Further, we describe a relationship to the recently highly popular dynamic mode decomposition (DMD) \cite{schmid2010dynamic,williams2015data,kutz2016dynamic,li2017extended,bollt2019geometric}, which is an empirical formulation of Koopman spectral theory, \cite{bollt2019geometric,arbabi2017ergodic,bollt2018matching}.   So while we do not offer this simplified  RC for performance over other approaches, we hope that  this work will serve to shed light on how the simplified RC approach is capable of providing useful time-series forecasts, and likewise as a suggestion as to how the general RC is successful.  {\color{black} There are related concepts concerning how an RNN is closely related to a NARMA model of a stochastic process (nonlinear autoregressive moving average), found in \cite{connor1992recurrent}. While this paper is mostly motivated to describe connections between different approaches, the machine learning RC approaches, the econometrics time-series VAR approach, and also the dynamical systems operator theoretic DMD approach, we show reasonable but not excellent forecasting ability of the linear RC with linear read-out equivalent of a VAR.  However, we do go on to connect a linear RC with quadratic read-out (as quadratic read-out popular for reasons of matching signal parity so it is described, \cite{pathak2018model}) which we show explicitly can be written as a quadratic NVAR.  That nonlinearities of the reservoir may be usefully moved to the output layer has been pointed out as a possibility and of practical use when building a photonic device implementation in \cite{vandoorne2014experimental}. 
 }

This paper is arranged as follows.  In Sec.~\ref{stochproc} we describe the nature of the data as derived from a stochastic process.  In Sec.~\ref{RCreview}, we review the standard RC concept, and we demonstrate it already with time-series data from a Mackey-Glass differential delay equation.  In Sec.~\ref{ReviewLinRC}, is the heart of this paper, where we first present that a linear activiation function allows the RC to be stated as a linear recursion, and therefore fitting just the read-out weights can proceed in a manner such that despite random read-in and inner weights, there is a well posed problem.  Then, in this form we are able to directly relate the linear RC solution to the classical VAR(k) solution.  As such, we are then able to enlist statistical time-series forecasting theory for forecasting stochastic processes, so that the Wold theorem that guarantees a VMA can then be translated to a VAR.  Furthermore the associated companion form of a VAR(1) usefully states the full vectorized problem.  In Sec.~\ref{IsKoop}, we note that the companion form of the VAR(1) is reminiscent of prior work for another famous concept in data-driven dynamical systems which is the time-delay formulation of a DMD-Koopman analysis.  In the examples Sec.~\ref{examples}, we present two classical examples, the Mackey-Glass differential delay equation and the Lorenz63 ordinary differential equation, with examples comparing aspects of a full nonlinear RC and the linear variant of an RC. {\color{black} We will consider the issue of fading memory in Sec.~\ref{vanishing}.  Finally, in Sec.~\ref{NLinRC}, show that a linear RC but with Hadamard quadratic readout is equivalent to a quadratic NVAR of all monomial quadratic terms, analogous to the earlier result of a VAR.}


\section{The  Data as Sampled From a Stochastic Process}\label{stochproc}

\begin{figure}[htbp]
\centering
\includegraphics[scale=.5]{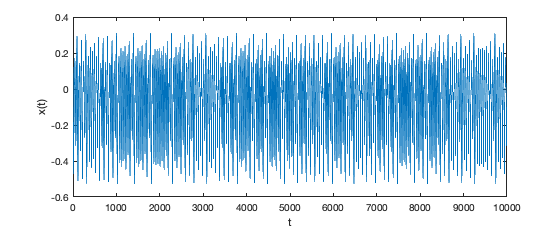}
\includegraphics[scale=.32]{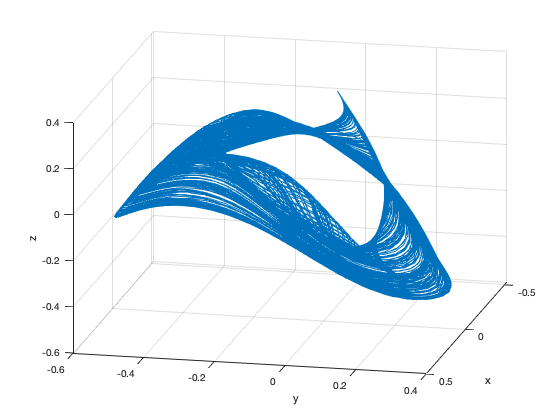}      
\caption{Time-series acquired from the Mackey-Glass differential delay equation, Eq.~(\ref{MGE}), has become a standard example for time-series forecasting, for benchmarking data-driven methods since it is dynamical rich and high-dimensional and therefore challenging.  (Top) Time-series, index.  (Bottom) Three-dimensional projection in delay coordinates, $(x(t),x(t-\tau), x(t-2*\tau))$, $\tau=20$. A sample of $N=10,000$ data points is chosen as the training data set.}
\label{fig:LSM1}
\end{figure}


For data-driven forecasting problems, we require data from a process,  including from a deterministic or otherwise from a stochastic dynamical system, \cite{bollt2013applied}.
A process, stated,
\begin{equation}\label{stochprocess}
    \{X_t:t\in T\}
\end{equation}
is in terms of a collection of random variables, $X_t$ on a common probability space, $(\Omega, {\cal B}, P)$, where $\Omega$ is the sample space, ${\cal B}$ the $\sigma$-alebra, and $P$ a corresponding probability measure.  $T$ is a ``time" index set and commonly it is chosen as either ${\mathbb R}$, or ${\mathbb Z}$ or subsets.  For sake of discussing finite samples of data, we emphasize maps, which may well be from discretely sampling a flow.  A data set from such a process samples $\bx_{t_i}$ of  $X_{t_i}$, stated as a time sorted sample, $\{\bx_i\}_{i=1}^N$, $t_1<t_2<...<t_N$, using  indexing notation, $\bx_i:= \bx_{t_i}$.  Uniform timing is also a simplifying assumption, $h=t_{i+1}-t_i$, for all $t_i\in T$.  Assuming a vector real valued time-series, of dimension $d_x$, $\{\bx_i\}_{i=1}^N\subset {\mathbb R}^{d_x}$. Data derived from
a flow, say,
\begin{equation}
\dot{\bx}={\mathbf f}(\bx)
\end{equation}
may be collected by stroboscopic map,
\begin{equation}\label{stochprocess2}
\bx_{i+1}={\mathbf F}_t(\bx_i)=\bx(t+\tau)=\bx(t)+\int_t^{t+\tau} {\mathbf f}(\bx(s)ds.
\end{equation}
Suppressing the stroboscopic time $t$, this is a discrete time map ${\mathbf F}$, and likewise other Poincare' maps may be useful for flight between surface of section, and random dynamical systems may also be relevant \cite{scott2006encyclopedia, bollt2013applied}.
%
An underlying principle here is that the data should be ``long enough", and likewise a general failing of any data-driven machine learning method for forecasting a stochastic process will tend to do much better in terms of interpolation than extrapolation.  Generalizing, to allow for out of sample forecasts will tend to fare much better when the point to be forecasts is  close to other observed inputs.  Said another way, the quality of results can be  brittle,  depending as much upon curating a representative data set as the details of the method used to avoid that struggle between fitting between observations and overfitting and too far out of sample. 

As a matter of presenting examples, we will highlight two classic problems that remain popular in benchmarking for  machine learning in recent literature.  These will be,
\begin{itemize}
 \item The Mackey-Glass differential delay equations, Eq.~(\ref{MGE}), and
\item  The Lorenz63 system, Eq.~(\ref{eq:lorenz}), 
 \end{itemize}
 both of which will be presented in fuller detail in Sec.~\ref{examples}.   In Fig.~\ref{fig:LSM1} we show early in this presentation for sake of context, a time-series data set of the Mackey-Glass system, from Eq.~(\ref{MGE}), to stand in as a typical data set.   This problem is a useful benchmark, and it is often used as such \cite{mackey1977oscillation,han2007reservoir,antonik2018using,ortin2015unified,bollt2000model, farzad2006predicting,yeo2017model}, perhaps because it is a well known chaotic process, but also for sake of dimensional complexities that we recall in Sec.~\ref{MGEsec}.


\begin{figure}[htbp]
\centering
\includegraphics[scale=.5]{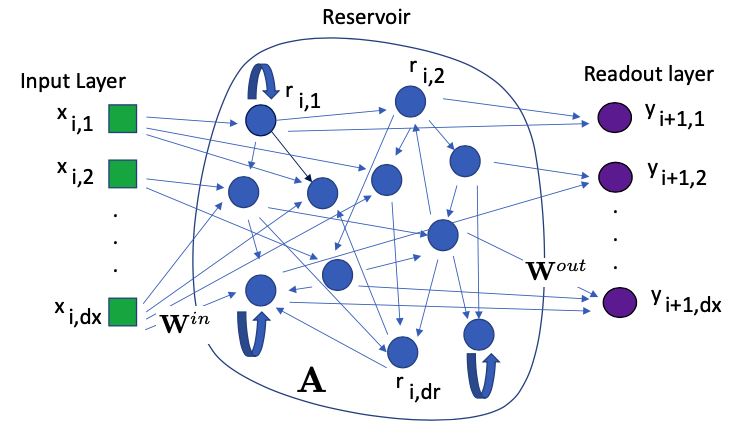}
\caption{Reservoir Computing (RC) as defined Eq.~(\ref{process1}), including a randomly selected $d_r\times d_x$ read in matrix, $\bW^{in}$ from $d_x\times 1$ states vector $\bx$, a randomly selected $d_r\times d_r$ inner layer recurrence matrix $A$ for inner states $d_r\times 1$ vector $\br$  and the $d_x\times d_r$ {\it trained read-out matrix} matrix $\bW^{out}$.}
\label{fig:rc}
\end{figure}

\section{Review of The Traditional RC  With Nonlinear Sigmoidal Activation Function}\label{RCreview}

In this section we review the standard and fully nonlinear RC method, by which we mean, including the use of a nonlinear activation function $q(s)$.  In this context, $q(s)$ is usually taken to be a sigmoidal function such as the hyperbolic tangent function.  However, in the next section we will challenge these steps including simplifying to the identity function, $q(s)=s$.

Assuming the training data, $\{\bx_i\}_{i=1}^N\subset {\mathbb R}^{d_x}$, the reservoir computing RNN is stated,
\begin{eqnarray}\label{process1}
    \br_{i+1}&=&(1-\alpha)\br_i+\alpha q({\mathbf A}\br_i+\bu_i+{\mathbf b}), \nonumber \\
    \by_{i+1}&=&\bW^{out} \br_{i+1}.
\end{eqnarray}
The hidden variable $\br_i \in{\mathbb R}^{d_r}$ is generally taken to be of a much higher dimension $d_r>d_x$, by a linear lifting  transformation, 
\begin{equation}\label{process2}
 \bu_i=\bW^{in} \bx_i,
\end{equation}
and $\bW^{in}$ is a randomly selected matrix $d_r\times d_x$ of weights.   See Fig.~\ref{fig:rc}.
${\mathbf A}$ is also a linear transformation, as randomly chosen square matrix $d_r\times d_r$ of weights, that should be designed with certain properties such as spectral radius for convergence \cite{carroll2019network, jiang2019model}, or sparsity, \cite{lu2018attractor, pathak2018model, vlachas2019forecasting} or otherwise consideration of the ``echo-state" property, \cite{buehner2006tighter}.  Likewise, the read-out is by a linear transformation, using a $d_x\times d_{r}$ matrix of weights $\bW^{out}$.  However, $\bW^{out}$, and only $\bW^{out}$,  is trained to the data, allowing for forecasts $\by_i$ given data $\bx_i\in {\mathbb R}^{d_x}$, which is the major simplify aspect of RC since it can be done by a simple and cheap least squares computation.  
Finally $q:{\mathbb R}\rightarrow {\mathbb R}$ is an ``activation" function, using the phrasing from machine learning in the neural network community to mimic the concept of a biological network that fires when a voltage has reached a threshold.  Popular choices include $q(s)=tanh(s)$, meaning a componentwise application of the scalar hyperbolic tangent function when $s$ is multivariate.  Other activations are popular in general neural network theory, including other sigmoidal functions, and also the ReLu function in certain contexts but not so commonly in RC, \cite{gauthier2018reservoir}.  $0\leq \alpha \leq 1$ serves to slow down the RC,   to moderate stability of the fitting, but we will restrict to $\alpha=1$ in this paper as outside the purpose of challenging the concept of explaining how the RC may work in a special case of identity $q$ in which case, nonzero $\alpha$ can be considered as absorbed into the random ${\mathbf A}$; $(1-\alpha)\br+\alpha {\mathbf A} \br=((1-\alpha)I+\alpha {\mathbf A} ) \br$ and since ${\mathbf A}$ is chosen randomly, then $((1-\alpha)I+\alpha {\mathbf A} ) $ may be an alternative random selection.  Finally, ${\mathbf b}$ serves as an offset for activation, that is useful in some contexts, but it is also not relevant for our needs for the same reason we choose $\alpha=1$, and we choose ${\mathbf b}=0$.

\begin{figure}[htbp]
\centering
\includegraphics[scale=.45]{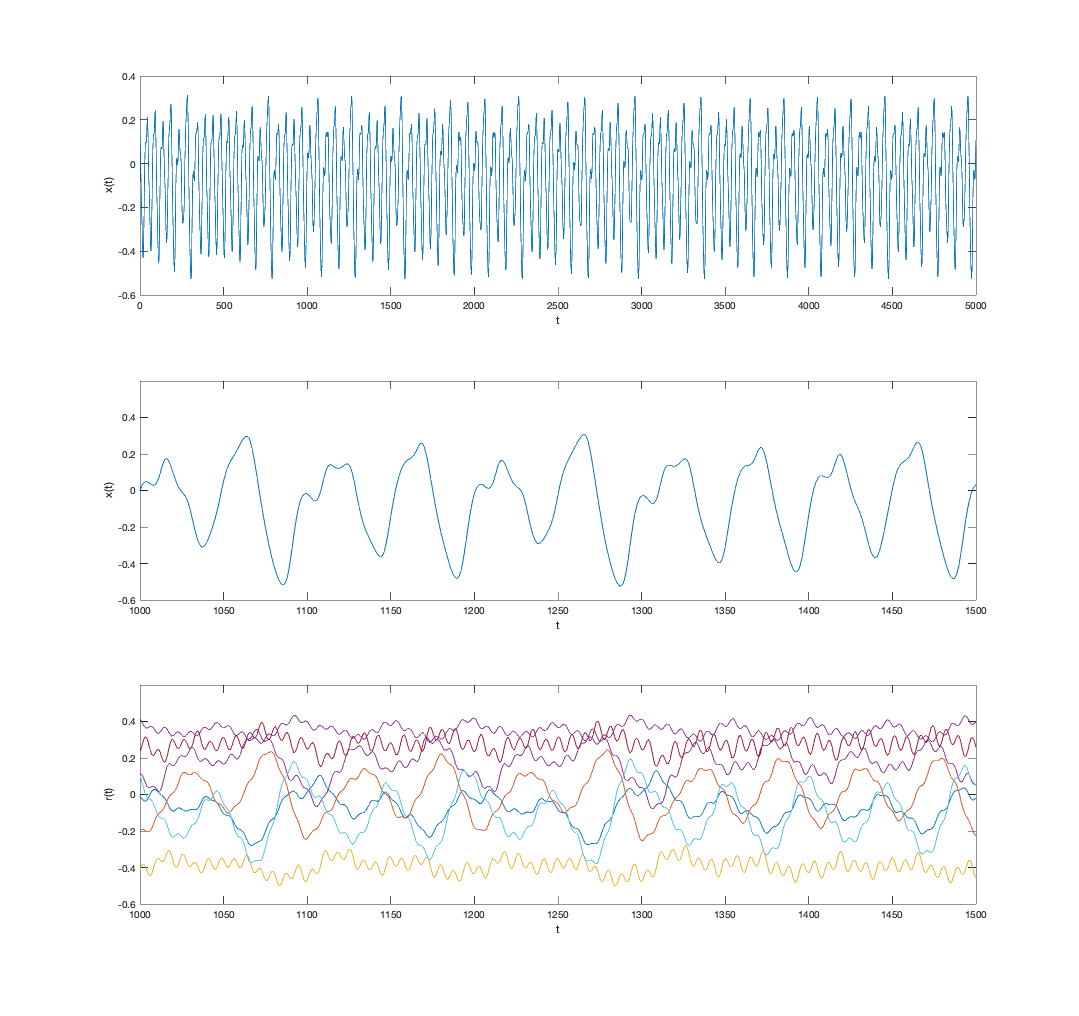}
\caption{Standard nonlinear RC, one-time-step forecasts from the Mackey-Glass differential delay equation, Eq.~(\ref{MGE}), using a training data set from $N=10,000$ samples as shown in Fig.~\ref{fig:LSM1}.  (Top) Time-series data, $N=5,000$ shown for clearer illustration. (Middle) Reservoir trained across the data set, and $500$ samples are shown for clarity where we see the error is sufficiently small that the one-time-step forecasts and the true data are almost the same so that the plot is indistinguishable (both shown, but curves overlay).  Regularity is chosen to be, $\lambda=1.e-6$.  (Bottom) Some randomly selected 7 of the (usually hidden) $d_r=500$ activation functions illustrate the general appearance.  Contrast to forecasting into the future as shown in Fig.~\ref{fig:LSM3}, and linear method in Fig.~\ref{fig:LSM4}.  }
\label{fig:LSM2}
\end{figure}

What is remarkable about RC is that the usual hard work of optimally developing a full RNN is almost entirely skipped.   Instead of learning $\bW^{in} $ and ${\mathbf A}$  optimally fitted to the data, these seemingly very important matrices are simply picked randomly.  This is an enormous savings over what would usually be inherently hard to handle since the parameters are composed within the nonlinear activation $q$ and require at least a gradient descent optimization of back-propagation in a high dimensional and likely multi-peaked optimization space.  Almost any matrix distribution may plausibly due, but several different recipes are suggested.
We say ``recipe" rather than algorithm since these are descriptions of successful observations in practice, rather than a product of mathematical theory that is still not complete. 
Here, we choose the entries of ${\mathbf A}$ uniformly, ${\mathbf A}_{i,j}\sim U(-\beta,\beta)$, with $\beta$ to scale the spectral radius, but other choices are common, notably for sparsity.  The read in matrix is also chosen uniformly randomly, $\bW^{in}_{i,j}\sim U(0,\gamma)$, with $\gamma>0$ chosen to scale the inner variables $\br$.  

The crucial aspect of the simplification that makes reservoir computing so easy and computationally efficient, is that training to the output becomes just a linear process.  The  cheap and simple least squares solution is easily handled directly by matrix computations.  Let,
\begin{equation}
    \bW_{out}=\argmin_{\bV\in{\mathbb R}^{d_{x}\times d_{r}}} \|\b X-\bV \bR\|_F=\argmin_{\bV\in{\mathbb R}^{d_{x}\times d_{r}}} \sum_{i=k}^N \|\bx_i- \bV \br_i\|_2, \mbox{ }k\geq 1.
\end{equation}
Notation here is  standard that $\|\cdot \|_F$ denotes the Frobenius-norm of the matrix, which is the least squares equivalent of the least squares matrix parameter estimation problem.
The data $\{\bx_i\}_{i=1}^N$  is stated  as a $d_{x}\times N-k$ array.
\begin{equation}
\bX=[\bx_{k+1} | \bx_{k+2} | \ldots | \bx_N]=[\bV \br_{k+1} | \bV \br_{k+2} | \ldots | \bV\br_N]=\bV \bR, \mbox{ }k\geq 1
\end{equation}
are the forecasts to $\bX$ to be optimized in least squares by $\bW^{out}$, processed through the RC, 
\begin{equation}\label{Rdata}
\bR=[\br_{k+1} | \br_{k+2} | \ldots | \br_N], \mbox{ }k\geq 1.
\end{equation}
While $k=1$ is allowable, here for theoretical development in subsequent sections, we  allow for larger $k\geq 1$, describing memory.  In practice a ridge regression (Tikhonov regularization with least squares regularity, \cite{gauthier2018reservoir, pathak2018model, golub2013matrix,bolltregularized}) is used to mitigate overfitting, the solution of which may be written formally,
\begin{equation}\label{RR}
    \bW^{out}:=\bX \bR^T(\bR \bR^T+\lambda {\mathbf I})^{-1}.
\end{equation}
Notation includes $\cdot^T$ is the matrix transpose, ${\mathbf I}$ is the identity matrix, and  the  choice of regularity parameter is $\lambda\geq 0$.  
We will write a regularized pseudo-inverse with the notation,
\begin{equation}\label{pseudo}
\bR^\dagger_\lambda:=\bR^T(\bR \bR^T+\lambda {\mathbf I})^{-1}
\end{equation}
 In Appendix \ref{appmtrix} we review the matrix theory as to how to form regularized pseudo-inverses such as $\bR^\dagger_\lambda$ by a regularized singular value decomposition (SVD) in terms of regularized singular values such as $\sigma_i/(\sigma_i^2+\lambda)$ obtained from the singular values $\sigma_i$ from the SVD of $\bR$. 
 
\begin{figure}[htbp]
\centering
\includegraphics[scale=.6]{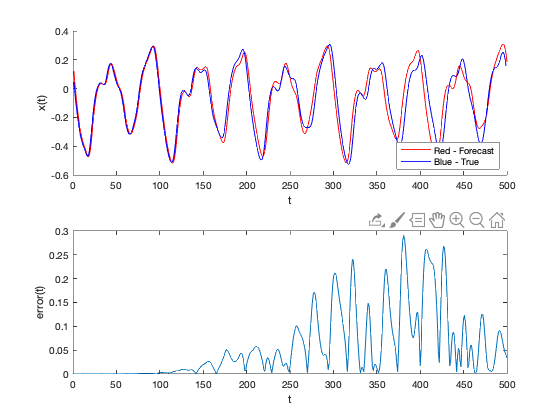}
\caption{Standard nonlinear RC, forecasts into the future, from the Mackey-Glass differential delay equation, Eq.~(\ref{MGE}), using a training data set from $N=10,000$ samples as shown in Figs.~\ref{fig:LSM1}, \ref{fig:LSM2}.  (Top) Time-series data, $0\leq t \leq 500$ zoom plotted for clearer illustration. (Top) Forecasts into the future (Red) diverge from true (Blue), and (Bottom) error is shown. {\color{black} All forecasts shown will be closed-loop, once trained on training set, style feeding RC output to input, but comparing for error to true evolution, as is standard such as \cite{lu2018attractor}. } }
\label{fig:LSM3}
\end{figure}

  In Fig.~\ref{fig:LSM2}, we show an example of an RC machine obtained from data obtained from the Mackey-Glass differential delay equations, Eq.~(\ref{MGE}).  We see fitting for $N=10,000$ data points $x(t)$, $d_x=1$, regularizing parameter $\lambda=1.0 \times 10^{-8}$, and fitting for constant time offset.  Fit and true data are shown to be so close that in fact the blue fit curve  hides the red true data curve.  Also shown are several (7) of the $d_r=500$ hidden variables $r(t)$.  The fit matrix ${\mathbf A}$ is randomly chosen with entries from a uniform distribution, and then scaled so that the spectral radius $\rho({\mathbf A})=1$.  The random random matrix $\bW_{in}$ is also chosen uniformly, scaled so that $x$ values lead to $\br$ in $[-0.6,0.6]$.  In Fig.~\ref{fig:LSM3},  the trained RC are used to forecast into the future.  We see small errors grow in scale, as illustrated by the bottom error curve.  Results from an RC forecasting for the Lorenz63 system are presented in Sec.~\ref{lorensec}, and notably the forecasting quality degrades more quickly in part due to known large Lyapunov instability of that system.

What is amazing is that despite that RC may  seemingly be a  gross oversimplification of the RNN concept, it still seems to work quite well. Also from experience, it is  generally stable in that it is somewhat insensitive to the parameters and hyperparameters of the fitting process, even if the level of quality does depend on these.  Furthermore, once it starts to make larger errors the kind of dynamics it produces are still plausible alternative wave forms of the process.  Nonetheless there are some parameter choices to make,   notably, $d_{r}>d_x$ must be ``large enough," but how big is not well understood.  Furthermore, the nature of the underlying distribution of  matrices $\bW^{in}$ and $\bA$ is not fully understood.  {\color{black} We hope to contribute some general perspective as to why an RC may work at all.  Our goal here is  not specifically to  improve performance, and admitting that the corresponding VAR makes reasonable short term forecasts, but perhaps no better than that, as illustrated in our examples.  However, we go on in  
Sec.~\ref{NLinRC}, with  details in Appendix \ref{nvarapp}, to show that fitting a quadratic read-out, that is extending Eq.~(\ref{Rdata}) to also include terms $\br \circ \br$ (componentwise multiplication, ``$\circ$" is called the Hadamard product) yields a quadratic NVAR of all monomial quadratic terms, which we observe performs quite well.}

\section{RC With A Fully Linear Activation, $q(s)=s$, Yields a VAR(k)}\label{ReviewLinRC}

Now we attempt to challenge a central typical assumption of the RC method.  Instead of choosing the activation function to be a sigmoid function, instead, we use the identity function, $q(x)=x$.  With this assumption, we can show that the resulting linear RC machine is equivalent to a vector autoregressive process (VAR)  \cite{qin2011rise, harrison2003multivariate}, which is extremely popular and successful in the timeseries forecasting field, particularly in econometrics  \cite{amisano2012topics}.  With this simplification, we find that not only can the linear RC still make useful forecasts, but we are able to connect the RC concept to this well established theory associated with VAR time-series analysis, notably the existence of representation WOLD theorem, \cite{wold1954study,pandit1983time}.  However, while this gives some explanation as to why a standard nonlinear RC may work despite the seemingly oversimplification of a full RNN, we show that that the linear RC does still performs and furthermore, now with theoretical underpinnings, even if the full nonlinear RC may still perform better.  So it is for the theoretical connections that we make this simplification, rather than a suggestion that it may be a new or  simpler method.

Before proceeding with a discussion of $q(s)=s$, notice that $r$ is related to the scale of the read-in matrix, $\bW^{in}$.
Proceed by initializing the process, by Eq.~(\ref{process2}),
\begin{equation}\label{zeroit}
    \bu_1=\bW^{in} \bx_1, \mbox{ but also we choose, } \br_1=0.
\end{equation}
Consider that since $\bW^{in}$ is randomly chosen, and we choose uniformly $\bW^{in}\sim U(0,\gamma)$, then the parameter $\gamma>0$ moderates the subsequent scale of terms $\bu_i$ and then $\br_i$.  See for example Fig.~\ref{fig:LSM2}, where the native data $\bx$ from the Mackey-Glass system is translated to scaled internal variables.  Recall the power series of the nonlinear activation function,
\begin{equation}
q(s)=tanh(s)\approx s-s^3/3+s^5/5-\ldots,
\end{equation}
Clearly for $s<<1$, then $q(s)\sim s$ even if chosen as a sigmoid, and the choice of read-in scale could be designed to put us in this regime as long as $\bA$ is designed to keep us in this regime.  
{\color{black}
That is, if we choose the scale of the read-in matrix, $0< \gamma <<1$, giving small values of the matrix $\bW^{out}_{i,j}$, then at least for a stable RC such as when $\bA$ has sufficiently small spectral radius, then  the arguments of $s$ from $\bA \br + \bu + {\mathbf b}$ in Eq.~(\ref{process1}) remain small.  So in practice $tanh(s)\sim s$.  Stated roughly, of $\gamma$ is small, then at least for some short time we might expect that the fully nonlinear RC is close to a fully linear RC. To advance beyond that as an idea, for now, we believe the insights gained for a linear activation RC should be relevant to the general problem.
} 

In the following we proceed to study the consequences of stating the  activation exactly as the identity, 
\begin{equation}
q(s)=s.
\end{equation}
With this assumption, the first several iterations follow from Eq.~(\ref{process1}) and Eq.~(\ref{zeroit}) as a forward propagation, for which we  explicitly observe the following recursion.
\begin{eqnarray}\label{iterate}
    \br_2&=&\bA \br_1+ \bu_1=\bu_1=\bW^{in} \bx_1 \\
    \br_3&=&\bA \br_2+ \bu_2 \nonumber \\
    &=& A \bW^{in} \bx_1+\bW^{in} \bx_2  \\
    \br_4&=&\bA \br_3+ \bu_3 \nonumber \\
    &=& \bA(\bA \br_2+\bu_2) +\bu_3 \nonumber \\
    &=&\bA^2 \bW^{in} \bx_1+\bA \bW^{in} \bx_2+ \bW^{in}\bx_3 \\
    &\vdots & \nonumber \\
    \br_{k+1}&=&\bA \br_k+ \bu_k \nonumber \\
    &=& \bA(\bA \br_{k-1}+\bu_{k-1}) +\bu_{k} \nonumber \\
    &\vdots& \nonumber \\
    &=&\bA^{k-1} \bW^{in} \bx_1+\bA^{k-2} \bW^{in} \bx_2 +\ldots + \bA \bW^{in} \bx_{k-1} +\bW^{in} \bx_{k}  \label{iteratee}\\
    &=&\sum_{j=1}^{k} \bA^{j-1} \bu_{k-j+1}=\sum_{j=1}^{k} \bA^{j-1} \bW^{in} \bx_{k-j+1}, \label{fin}
\end{eqnarray}
using notation, $\bA^0=I$, the identity matrix.
Since the read-out of this process is by Eq.~(\ref{process1}), 
$\by_i=\bW^{out} \br_i$, then we may rewrite the final equation, Eq.~(\ref{fin}), by left multiplying by $\bW^{out}$.
\begin{eqnarray}\label{arnoldi1}   
    \by_{\ell+1}&=& \bW^{out} \br_{\ell+1} \nonumber \\
    &=& \bW^{out}  \sum_{j=1}^{\ell} \bA^{j-1} \bW^{in} \bx_{\ell-j+1} \nonumber \\
    &=&\bW^{out} \bA^{\ell-1} \bW^{in} \bx_1+\bW^{out}\bA^{\ell-2} \bW^{in} \bx_2 +\ldots + \bW^{out}\bA \bW^{in} \bx_{\ell-1} +\bW^{out} \bW^{in} \bx_{\ell} \nonumber \\
    &=& a_\ell\bx_1+a_{\ell-1}\bx_2 +\ldots + a_{2} \bx_{\ell-1} +a_{1}\bx_{\ell}, \label{VAR1}
\end{eqnarray}
with notation,
\begin{equation}\label{prod1}
    a_{j}=\bW^{out} \bA^{j-1}\bW^{in}, \mbox{ } j=1, 2,...,\ell.
\end{equation}
Each of these coefficients $a_j$ are $d_x\times d_x$ matrices.  This follows simply by Eq.~(\ref{prod1}), collecting products between $d_x \times d_r$ to $d_r\times d_r$ and then $d_r\times d_x$ matrices and   notation $\bA^{l}=\Pi_{i=1}^l \bA =\bA \cdot \bA \ldots \cdot \bA$, $l$-times if $l>0$, or the identity matrix when $l=0$.  

{\color{black} In some sense, Eq.~(\ref{prod1}), and quadratic generalization Eqs.(\ref{prod11})-(\ref{prod3}), are the heart of this paper as it is an explicit representation of the coefficient matrices of a VAR, (or NVAR), but as found in terms of projection onto iterations of the  random matrices involved in developing a linear activation function version of an RC.   With exactly $k$ VAR matrices $a_j$, the randomness of the $d_r^2$ free parameters of the random matrix $\bA$ collapses onto $k d_x^2$ parameters, meaning it yields only the finitely many fitted parameters of the matrices of $a_1, ... a_k$.  However, for a longer time observations which is the more usual way an RC is trained in practiced, Eq.~(\ref{prod1}) implies that when condition,
\begin{equation}\label{condition}
d_r<k d_x,
\end{equation}
 then randomness of the choice of $\bA$ and $\bW^{in}$ is completely specified by stating matrices $\bA^{j-1} \bW^{in}$, for many $j$.    We will expand upon this statement in the next section, and then how it relates to vanishing memory in Sec.~\ref{vanishing}.
 }

By Eq.~(\ref{VAR1}),  a linear RC yields a classical  VAR(k), (a vector autoregression model of $k$-delays) that in a general form  is  \cite{qin2011rise},
\begin{equation}\label{vark1}
    \by_{k+1}=c+ a_k\bx_1+a_{k-1}\bx_2 +\ldots + a_{2} \bx_{k-1} +a_{1}\bx_{k}+\boldsymbol\xi_{k+1}.
\end{equation}
In this writing,  $c$ allows for a general offset term, a $d_x\times 1$ vector that here we do not pursue.  The  $\boldsymbol\xi_{k+1}$ is underlying ``noise" of the stochastic process which is part of the stability theory we review in the next section, must be assumed to come from a covariance stationary process.  This relationship between an RC and a VAR(k)  allows us to relate to  the corresponding theoretical discussions of relevant alternative forms and stability and convergence from the stochastic process time-series literature, that we will also expand upon in the next section.  


Considering the complete data set of vector time-series, $\{\bx_i\}_{i=1}^{N}$ yields,
\begin{equation}\label{fit}
\begin{bmatrix}
| & | & | & | \\
\by_{k+1} & 
\by_{k+2} & 
\ldots &
\by_{N} \\
| & | & | & |
\end{bmatrix}=
\begin{bmatrix}
\begin{bmatrix}
a_1
\end{bmatrix}
&
\begin{bmatrix}
a_{2}
\end{bmatrix}
&
\ldots &
\begin{bmatrix}
a_k
\end{bmatrix}
\end{bmatrix}
\begin{bmatrix}
 |& | & \vdots &  |\\ 
\bx_k& \bx_{k+1} & \ldots & \bx_{N-1}  \\ 
|& | & \vdots &  |\\ 
 \bx_{k-1}& \bx_k & \ldots &  \bx_{N-2} \\ 
|& | &\vdots & |\\ 
\vdots & \vdots & \vdots & \vdots \\
|& | &\vdots & |\\ 
\bx_{1} & \bx_{2} & \ldots &  \bx_{N-k}\\
|& | & \vdots &  |\\ 
\end{bmatrix}.
\end{equation}

Restating this as a single linear equation,  
\begin{equation}\label{meq}
   \bY=\ba {\mathbb X}.
\end{equation}
Again, remembering that $\bx_i$ are $d_x\times 1$ vectors and that $a_i$ are $d_x\times d_x$ matrices, 
$\ba=[\begin{bmatrix}
a_1
\end{bmatrix}
|\begin{bmatrix}
a_{2}
\end{bmatrix}
|\ldots | \begin{bmatrix}
a_k
\end{bmatrix}
],
$ is a $d_x \times (k d_x)$ matrix. 
$ \bY=\begin{bmatrix}
\by_{k+1} |
\by_{k+2} | 
\ldots |
\by_{N}
\end{bmatrix},
$ is a $d_x\times (N-k)$ matrix, and $ {\mathbb X}$ is a $(k dx) \times (N-k) $ matrix.  

Formally, minimizing in least squares, with regularization,
  \begin{equation}\label{ls1}
     J(\ba)=\| \bY-\ba  {\mathbb X}\|_F+ \lambda \| \ba \|_F,
 \end{equation}
 with $\bY$ being the target output of the right hand side of Eq.~(\ref{fit}) by best fitted matrix $\ba^*$.
The solution of this regularized least squares problem may be written in its matrix form,  
\begin{equation}\label{ls2}
    \ba^*=\bY {\mathbb X}^T( {\mathbb X} {\mathbb X}^T+\lambda I)^{-1}:=\bY  {\mathbb X}_\lambda^\dagger, 
\end{equation}
where the symbol $^\dagger$ refers to the Penrose pseudo-inverse, with notation described in detail  in Eqs.~(\ref{tik1})-(\ref{tik2}),  when formulating the ``ridge" Tikhonov regularized pseudo-inverse $ {\mathbb X}^\dagger_\lambda$.

\begin{figure}[htbp]
\centering
\includegraphics[scale=.55]{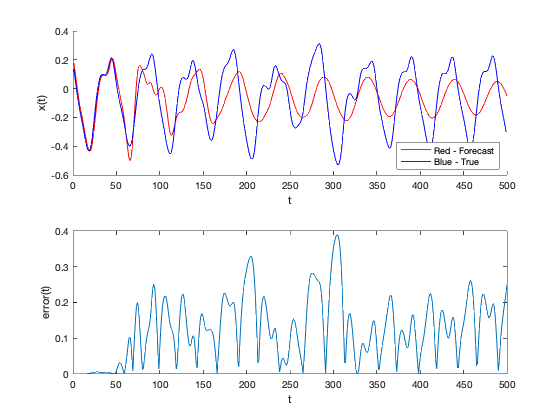}
\caption{The fully linear RC, $q(s)=s$,   forecasts from the Mackey-Glass differential delay equation, Eq.~(\ref{MGE}), using the same training data set from $N=10,000$ samples as shown in Fig.~\ref{fig:LSM1}.  Contrasting to forecasts into the future as shown in Figs.~\ref{fig:LSM3}, we see that clearly the nonlinear RC outperforms the linear RC, and by a wide margin.  But that is not the message here, rather which is one of explaining the relationships and fitting of the parameters, and so that fitting just the read-out matrix $\bW^{out}$ is relevant is established by Eq.~(\ref{ls4}).  \label{fig:LSM4}}
\end{figure}

\subsection{Decomposing the VAR(k) Solution Explicitly Relates to RC}


Now, we will further decompose the derived VAR(k) coefficients found in  Eq.~(\ref{ls2}), to emphasize the training of just the output matrix $\bW_{out}$ of an associated RC, in terms of randomly pre-choosing $\bA$ and $\bW^{in}$.

Referring to Eqs.~(\ref{arnoldi1})-(\ref{prod1}), we can rewrite Eqs.~(\ref{fit})-(\ref{meq}) as,
\begin{equation}\label{collapse1}
\bY=\ba  {\mathbb X} = \bv  {\mathbb A} {\mathbb X}.
\end{equation}
with the matrix defined,
\begin{equation}\label{Akry}
 {\mathbb A}=
[
 \bW^{in} | \bA \bW^{in} | \ldots |\bA^{k-2} \bW^{in} | \bA^{k-1} \bW^{in} 
],
\end{equation}
This ${\mathbb A}$ is a combination of exponents of the random $d_r\times d_r$ matrix $\bA$, and the random $d_r\times d_x$ matrix $\bW^{in}$, and so it is itself a $d_r\times k d_x$ random matrix.  
Interestingly, considering just one column at a time of the $\bW^{in}_l$, $l=1,2..,d_r$, $\bA^{k-1}$ can be understood as a collection of columns from a Krylov space and this entire process can be discussed as an Arnoldi-iteration, which is something we will explore further in Section \ref{IsKoop}.

Consider that the least squares objective Eq.~(\ref{ls1}) can be expanded to split,
\begin{equation}\label{f1}
\ba=\bv  {\mathbb A},
\end{equation}
 to emphasize that since if we pre-choose $\bA$ and $\bW^{in}$, then only the read-out matrix $\bv$  is a free parameter,
\begin{equation}\label{ls1}
     J(\bv)=\| \bY-\ba  {\mathbb X}\|_F=\| \bY-\bv  {\mathbb A}  {\mathbb X}\|_F.
 \end{equation}
 Optimizing for $\bv$ yields,
 \begin{equation}\label{ls3}
 \bW^{out}:=\bv^*=\bY( {\mathbb A}  {\mathbb X})^\dagger=(\bY  {\mathbb X}^\dagger)  {\mathbb A}^\dagger,
 \end{equation}
Comparing this equation with Eq.~(\ref{ls2}), defining $\ba$, we see $(\bX  {\mathbb X}^\dagger) $  formally appears in both expressions. Only  the associative property of matrix multiplication is needed to emphasize the role of $ {\mathbb A}$.    More importantly, this expression Eq.~(\ref{ls3}) for   $\bW^{out}$ is written so as to emphasize that the reservoir computing process is designed with $ {\mathbb A}$ and $ {\mathbb X}$.  Combined through the iteration, as $( {\mathbb A}  {\mathbb X})$ is the data that results from Eq.~(\ref{iteratee}),
\begin{equation}
\br_{\ell+1}=A^{\ell-1} \bW^{in} \bx_1+A^{\ell-2} \bW^{in} \bx_2 +\ldots + A \bW^{in} \bx_{\ell-1} +\bW^{in} \bx_{\ell} .
\end{equation}
This is written naturally,
\begin{equation}
\bR=({\mathbb A} {\mathbb X}).
\end{equation}
by the simple way Eq.~(\ref{ls3}) uses a matrix identity of pseudo-inverses, \cite{golub2013matrix}, 
\begin{equation}
({\mathbb A} {\mathbb X})^\dagger={\mathbb X}^\dagger{\mathbb A}^\dagger.
\end{equation}
 Associativity  emphasizes that since ${\mathbb A}$ is deterministically defined, once $\bA$ and $\bW^{in}$ are chosen, and separately from the data $\bX$, then the fitting of only the parameters of $\bW^{out}$ are sufficient.  If we  want the VAR(k) parameters, we could either ignore the  prior knowledge of choice of $\bA$ and $\bW^{in}$, and compute $\ba$ directly from Eq.~(\ref{ls1}), or from Eq.~(\ref{ls3}), defining,
\begin{equation}\label{ls4}
\bW^{out}:=\bv^*=\ba^* {\mathbb A}_\lambda^\dagger=\bY {\mathbb X}_\lambda^\dagger {\mathbb A}^\dagger_\lambda.
\end{equation}

We summarize that these manipulations concluding with Eq.~(\ref{ls4})
serve directly as the connection between the RC fitted read-out and the coefficient matrices of a VAR(k).  The roles of pre-choosing $\bA$ and $\bW^{in}$ relate directly to $\bW^{out}$ coefficients, or indirectly to the fitted data.  
{\color{black} 
Considering the training of $\bW^{out}$, Eqs.~(\ref{f1}), (\ref{ls3}), (\ref{ls4}), in terms of  geometric description of least squares estimation \cite{golub2013matrix},  $\bW^{out}$best estimates  orthogonal projections of rows of $\ba$ into the row space of ${\mathbb A}$ which has $d_r$-row vectors of dimension $k d_x$.  So no more than $d_r$ dimensions can remain free, or as described similarly in inequality Eq.~(\ref{condition}).
}

Concluding this section with the an example, we simplify the nonlinear RC of the Mackey-Glass data from Figs.~\ref{fig:LSM1}, \ref{fig:LSM2}, to a purely linear RC fit shown in Fig.~\ref{fig:LSM4} which clearly is not as well performing but it does still make some forecast into the future.   Further discussion of this example and also likewise a Lorenz63 example in Sec.~\ref{examples}.  Said similarly, ${\mathbb A}$ has no more than rank $dr$

\section{VAR(k) Theory Suggests Convergence with $k$}\label{generalvar}

Since the VAR(k) model of vector autoregression appears naturally in our discussion from the simplified activation function $q(x)=x$,  as summarized by Eqs.~(\ref{arnoldi1}), and (\ref{vark1}), we now recall  some of the classical underlying theory from  the statistical time-series analysis literature \cite{wold1954study, qin2011rise} that describes sufficient conditions under which we expect existence of a VAR(k) representation.  

The Wold theorem plays a central role in time-series analysis as it describes existence of a vector moving average (VMA) model representation, which then under further assumptions for invertibility, is equivalent to a VAR.
Assumptions require a stationary process as a sum of two components: 1) a stochastic component consisting of ``linear" combinations of lags from a white noise process, and 2) a deterministic component that is uncorrelated with the stochastic component.   First we recall definitions.  A d-dimensional stochastic process $\boldsymbol\xi_t$ of zero mean, ${\mathbb E}(\boldsymbol\xi_t)={\mathbf 0}$,
is derived from a
white noise stochastic process, written 
with zero mean $ \boldsymbol\xi_t=[\xi_{1,t},\xi_{2,t},...,\xi_{d,t}]\sim WN(0,\Omega)$  if ${\mathbb E}(\boldsymbol\xi_t)={\mathbf 0}$ and ${\mathbb E}(\boldsymbol\xi_{t_1}\boldsymbol\xi_{t_2}^T)={\mathbf 0}$,   for $t_1\neq t_2$, but 
${\mathbb E}(\boldsymbol\xi_{t}\boldsymbol\xi_{t}^T)=\Omega$
 is symmetric positive semi-definite.   A stochastic process is covariance stationary if all terms of the sequence have the same mean, and any two terms depend only on their relative positions.  That is, ${\mathbb E}( \boldsymbol\xi_{t'})={\mathbb E}( \boldsymbol\xi_{t})$,
  for all $t'$, and for all $t'\geq 0$, there exists $\gamma_{t'}\in {\mathbb R}$ such that, $Cov( \boldsymbol\xi_t, \boldsymbol\xi_{t-t'})=\gamma_{t'}$, for all $t>t'$, meaning depending on $t-t'$ rather than the $t$ or $t'$.  With these definitions, we can state the central theorem of this section that we recall:

\begin{theorem}[Wold Decomposititon Theorem, \cite{wold1954study, qin2011rise}]\label{woldthem} 
A zero mean covariance stationary vector process $\{\bx_t\}$ admits a representation,
\begin{equation}
\bX_t=C(L) \boldsymbol\xi_t+\boldsymbol\mu_t, 
\end{equation}
where $C(L) =\sum_{i=0}^\infty C_i L^i$ is a polynomial delay operator polynomial, the $C_i$ are the moving average matrices, and $L^i(\boldsymbol\xi_t)= \boldsymbol\xi_{t-i}$.  The term $C(L)\boldsymbol\xi$ is the stochastic part of the decomposition.  The $\boldsymbol\mu_t$ term is the deterministic (perfectly predictable) part as a linear combination of the past values of $\bX_t$.
Furthermore, 
\begin{itemize}
\item $\boldsymbol\mu_t$ is a $d$-dimensional linearly deterministic process.
\item $\boldsymbol\xi_t\sim WN(0,\Omega)$  is white noise.
\item Coefficient matrices are square summable, 
\begin{equation}\label{squaresum}
	\sum_{i=0}^\infty \|C_i\|^2<\infty.
\end{equation}
\item $C_0=I$, the identity matrix.
\item For each $t$, $\boldsymbol\mu_t$ is called the innovation or the linear forecast errors.
\end{itemize}
\end{theorem}
Clarifying notation of the delay operator polynomial, with an example, let, 
\begin{equation}
C(L)=\begin{bmatrix} 1 & 1+L \\ -\frac{1}{2}L & \frac{1}{2} - L \end{bmatrix} = \begin{bmatrix}1 & 1\\0 & \frac{1}{2}\end{bmatrix}+\begin{bmatrix} 0 &1 \\ -\frac{1}{2} &-1 \end{bmatrix}L=C_0+C_1 L, \mbox{ and } C_i=
\begin{bmatrix}0 & 0 \\ 0 & 0\end{bmatrix} \mbox{ if } i>1, 
\end{equation}
so if for example, $\bx_t\in {\mathbb R}^2$,
\begin{equation}
C(L)\bx_t=\begin{bmatrix} 1 & 1+L \\ -\frac{1}{2}L & \frac{1}{2} - L \end{bmatrix} \begin{bmatrix}x_{1,t} \\ x_{2,t}\end{bmatrix}=\begin{bmatrix}x_{1,t}+x_{2,t}+x_{2,(t-1)}\\ \frac{1}{2}x_{1,(t-1)}+\frac{1}{2} x_{2,t}-x_{2,(t-1)}\end{bmatrix}.
\end{equation}

For  interpretation and definition, consider:
\begin{itemize}
\item If $\boldsymbol\mu_t=0$, then this is called a ``regular" process, and therefore there is a purely vector moving average (VMA) representation.  If $C_i=0$ for $i>p$ for some finite $p>0$ then it is called a VMA(p) or otherwise it is a VMA($\infty$) representation.
\item If $\bX_t$ is regular then the representation is unique.
\end{itemize}

Now to our point to relate a Wold VMA representation to our discussion following the linear RC where we saw a VAR(k) results Eqs.~(\ref{VAR1}) when the activation is linear $q(x)=x$.  If the delay polynomial $C(L)$ is invertible with, $C(L)^{-1}C(L)=I$, and denote $C(L)^{-1}=B(L)=B(L)=B_0-B_1L-B_2L^2-\ldots$ in terms of matrices $B_i$, then writing explicitly,
\begin{equation}\label{terms}
(B_0-B_1L-B_2L^2-\ldots)(I+C_1L+C_2L^2+\ldots)=I.
\end{equation}
{\color{black} Existence of this inverse implies that the Wold implied VMA process has a representation,
\begin{equation}\label{varvma}
\bX_t=C(L) \boldsymbol\xi_t \implies B(L) \bX_t= \boldsymbol\xi_t,
\end{equation}
that is a VAR representation in that this represents the latest $\bX_t$ as a linear combination of prior values of $\bX$ written succinctly in terms of the delay operator $B(L)$.}


In practice, when an infinite order vector moving average process, VMA($\infty$) corresponds to an infinite order vector autoregressive process, VAR($\infty$), then recursion of expanding Eq.~(\ref{terms}) and matching term by term yields, 
\begin{equation}\label{terms2}
B_0=I, B_1=C_1, \ldots, B_k=C_k+B_1C_{k-1}+\ldots +B_{k-1}C_1,...
\end{equation}
 Though, a VAR representation may be found from a VMA through several methods, including a method of moments leading to the Walker-Yule equations, \cite{paulsen1985estimation}, or a least squares method in the case of finite presentations.  Often, for parsimonious efficiency of presentation, a mixed form of a $p$-step AR and a $q$-step MA model might make a suitable approximation, for what is called a ARMA(p,q) model.  
  

While not allowing ourselves to be drawn entirely into the detailed theory of econometrics and statistical time-series analysis, pursuing stronger necessary conditions, we wish to point out some already apparent relevant points  from the stated special sufficient conditions.
\begin{remark}
Summary statements.  If a vector stochastic process satisfies the hypothesis of a Wold theorem, then it:
\begin{itemize}
\item Can be written either as a VMA or a VAR, when Eq.~(\ref{varvma}) of $C(L)$ is invertible, Eq.(\ref{terms2}).
\item In practice a finite k, VAR(k) estimates a VAR($\infty$) as $k\uparrow$, since the sequence of coefficients matrices $\{C_i\}$, are square summable, Eq.~(\ref{squaresum}), and considering Eq.~(\ref{terms2}).
\item Furthermore, in practice a least squares estimate of a VAR(k) may be used for finite k, which relates  to an RC by the least squares fit, Eqs.~(\ref{ls3}), (\ref{ls4}).
\end{itemize}
\end{remark}

Finally, we separate from the above technical points, the following fundamental remark to distinguish existence versus uniqueness of a representation,
\begin{remark}
While a stochastic process may have a VMA representation and if through invertibility, a corresponding VAR, which is a linear descriptions of the process, it may not taken to be ``the" unique physical underlying description since  nonlinear descriptions   certainly may exist.  
\end{remark}

\begin{remark}
The processes that we may be interested in, such as those derived from Eq.~(\ref{stochprocess}), may describe the evolution of a (chaotic) dynamical system and these may allow a representation, Eq.~(\ref{stochprocess2}), \cite{bollt2013applied, sun2015causal,bollt2002manifold, lasota2013chaos}.  However, in many of these natural examples, the ``color" or even the nature of the noise may well not be conforming to the white noise assumption of the Wold theorem \ref{woldthem}.  Certainly contrasting samples from an invariant measure from a chaotic dynamical system to a white noise process is a well studied \cite{rosso2007distinguishing, kennel1992method}, but still undecided topic.  While existence of the VMA and corresponding VAR representation by referring to the Wold theorem does depend on that hypothesis, nonetheless, successful constructive  fitting of a VAR(k) by regression, even if implicitly through an RC, seems to proceed successfully in practice in a wide array of examples.
\end{remark}
With this last remark, we admit that while the details of the rigor guaranteeing existence may in practice break down, due to inability to check all hypothesis, as often such gaps occur between mathematics, applied mathematics, and practice as related to real world data, we feel that the concept is still highly instructive as underlying explanation, despite strong sufficient assumptions used to extend a rigorous theory.

{\color{black} We summarize this section that the relationship between the WOLD theorem for the VAR to our interest in an RC gives two conclusions.  1) Existence of the VMA representation follows the WOLD which in turn leads to a VAR when the delay operator is invertible.  2) That the coefficient matrices are square summable serves as an upper bound that memory must be fading.  We describe memory further in Sec.~\ref{vanishing}. }

\subsection{Stability of the VAR(k) and relationship to a VAR(1)}

To discuss stability, we recall \cite{qin2011rise} the fact that a VAR(k), Eqs.~(\ref{arnoldi1}),   (\ref{vark1}),
\begin{equation}\label{genvarp}
\bx_{k+1}=\bc+ a_k\bx_1+a_{k-1}\bx_2 +\ldots + a_{2} \bx_{k-1} +a_{1}\bx_{k}+\boldsymbol\xi_{k+1},
\end{equation}
can be stated as a VAR(1) in terms of ``stacked" (delayed) variables, called the companion system.  This idea is familiar in dynamical systems as we see it is related to stating time-delay variables and the Taken's embedding theorem \cite{takens1981detecting,packard1980geometry,sauer1991embedology,muldoon1998delay,bollt2000model,yonemoto2001estimating}. 
Define,
\begin{equation}\label{var1form}
{\mathbb X}_{k+1}={\cal A} {\mathbb X}_k + {\cal C} + \be_k, \mbox{ where, }
{\mathbb X}_k=
\begin{bmatrix}| \\ \bx_k \\ | \\ \bx_{k-1} \\ | \\ \vdots \\ | \\  \bx_1 \\ | 
\end{bmatrix}, \mbox{ }
\be_k= 
\begin{bmatrix}
\boldsymbol | \\ \xi_{k} \\ | \\
0 \\ | \\
\vdots \\
| \\ 0 \\ | 
\end{bmatrix},
{\cal C}=
\begin{bmatrix}
| \\ \bc \\ | \\ 0 \\ | \\ \vdots \\ | \\ 0 \\ |
\end{bmatrix}
 \mbox{ and }
{\cal A} =
\begin{bmatrix}
a_1 & a_2 & \ldots & a_k \\
I & 0 & \ldots & 0\\
0 & I & \ddots & 0 \\
0 & \ldots & I & 0
\end{bmatrix},
\end{equation}
where since $a_i$ are each $d_x\times d_x$ matrices,then ${\cal A}$ is $k d_x\times k d_x$, and ${\mathbb X}_k$ is $k d_x\times 1$.  For discussion in the next section, it will be convenient to consider for contrast  to Eq.~(\ref{fit2}), a matrix of all the data,
\begin{equation}\label{fit5}
{\mathbb X}=
\begin{bmatrix}
{\mathbb X}_k & {\mathbb X_{k+1}} & \ldots & {\mathbb X}_{N-k-1}
\end{bmatrix}, \mbox{ and likewise let, }
{\mathbb X}'=
\begin{bmatrix}
{\mathbb X}_{k+1} & {\mathbb X_{k+2}} & \ldots & {\mathbb X}_{N-k}
\end{bmatrix},
\end{equation}
are $k d_x \times (N-k-1)$.
Notice that the data in ${\mathbb X}$, also from Eq.~(\ref{fit}).

It follows that analysis of stability of a VAR(1) sufficiently describes the stability of a VAR(k).  If there is even a small offset $\bc$, whether by a bias or imperfection of fit,  then  follows the recursion,
\begin{equation}\label{var1}
{\mathbb X}_{k}={\cal C} + {\cal A}  {\mathbb X}_{k-1} + \be_{k-1}, 
\implies 
{\mathbb X}_{k}=(I+{\cal A} +\ldots +{\cal A} ^{l-1}) {\cal C} + A^l {\mathbb X}_{k-l} + \sum_{j=0}^{l-1} {\cal A} ^j  \be_{k-l}.
\end{equation}
This relates the VAR(1) back to a VMA($l$) form.
Clearly, even a small constant disturbance ${\cal C}$ is successively influenced by the (delay) matrix ${\cal A} $. 
In the limit $l\rightarrow \infty$, recall the geometric series of matrices,
\begin{equation}
(I-{\cal A} )^{-1}=\lim_{l\rightarrow \infty} (I+{\cal A} +\ldots +{\cal A} ^{l-1}),
\end{equation}
converges if the spectral radius is strictly contained in the complex unit disc,
\begin{equation}
\rho({\mathcal A})=\max_{\lambda: det(A-\lambda I)=0} |\lambda|<1.
\end{equation}
  Equivalently,  a general 
VAR(k), Eq.~(\ref{genvarp}), is stable if and only if a characteristic polynomial, 
\begin{equation}
det(I-a_1 z - a_2 z^2 - \ldots - a_k z^k )= 0,
\end{equation}
has all its roots outside the unit disc.
Under this stability  assumption we conclude that,
\begin{equation}\label{arnoldi4}
{\mathbb X}_{k}={\cal C} + {\cal A}  {\mathbb X}_{k-1} + \be_{k-1}= (I-{\cal A} )^{-1} {\cal C}  +  \sum_{j=0}^{\infty} {\cal A}^j  \be_{k-l}.
\end{equation}
which relates a VAR(1) form to a Wold form through a VMA($\infty$).
Since by Eq.~(\ref{prod1}), each matrix $ a_{j}=\bW^{out} \bA^{j-1}\bW^{in}$, then the magnitude of entries in the matrix $\bA$ and the read in matrix $\bW^{in}$ each moderate the magnitudes of entries of $ a_{j}$.  So considerations by the Gershgorin disc  theorem, \cite{golub2013matrix} relates these magnitudes to the magnitudes of $z$.  Generally sparsity of $\bA$, magnitude of the spectrum of $\bA$ and magnitudes of $\bW^{in}$ can be reduced for stability, and to moderate the ``memory" associated with converges with $k$, and these magnitudes were already discussed for sake of a regime where the usual sigmoidal $q$ would be close to the identity.

{\color{black}
\section{Quadratic Nonlinear VAR}\label{NLinRC}  

In this section, as is common practice, \cite{pathak2018model}, we investigate an RC that fits the readout $\bW^{out}$ using not just linear $\br$ values from the RC, but also terms $\br \circ \br$. Notation ``$\circ$" is the Hadamard product, meaning componentwise multiplication, $[\br \circ \br]_j=[\br]_j^2$ for each $j$.  This yields $\bW^{out}$ that is $d_x\times 2d_r$.  Here we state results briefly with detailed derivations given in Appendix \ref{nvarapp}.

Generalizing the VAR results of Sec.~\ref{ReviewLinRC},  here we state that a linear VAR with a Hadamard product-quadratic read-out is equivalent to a quadratic nonlinear VAR, (NVAR) with all quadratic terms in $r_ir_j$.  That is,
analogously to the VAR stated Eqs.~(\ref{arnoldi1}), (\ref{vark1}),
a quadratic nonlinear VAR may be stated, (abbreviated restatement of Eqs.~(\ref{NVAR1})-(\ref{prod2}), derived in Appendix \ref{nvarapp}),
\begin{eqnarray}\label{arnoldiq} 
    \by_{k+1}&=& a_k\bx_1+a_{k-1}\bx_2 +\ldots + a_{2} \bx_{k-1} +a_{1}\bx_{k}+ \nonumber \\
    & & +    a_{2,(k,k)} p_2(\bx_1,\bx_1)+   a_{2,(k-1,k)} p_2(\bx_2,\bx_1)+   ...+ a_{2,(1,1)} p_2(\bx_k,\bx_k)        \label{NVAR1}   
\end{eqnarray}
with notation for the $k$ linear coefficient $d_x\times d_x$ matrices,
\begin{equation}\label{prod11}
    a_{j}=\bW^{out}_1 \bA^{j-1}\bW^{in}, \mbox{ } j=1, 2,...,k.
\end{equation}
Now we have $k^2$ quadratic term $d_x\times d_x^2$  coefficient matrices, 
\begin{equation}\label{prod3}
a_{2,(i,j)}=\bW^{out}_2 P_2(A^{i-1} \bW^{in}, A^{j-1} \bW^{in}), i,j=1...k.
\end{equation}
The notation, $p_2(\bv,\bw)=[v_1w_1|v_1w_2|...|v_nw_n]^T$ defines a $n^2$-vector of all quadratic terms stated between vectors $\bv=[v_1|...|v_n]^T$, and $\bw=[w_1|..|w_n]^T$. $P_2(A^{i-1} \bW^{in}, A^{j-1} \bW^{in})$ is a $d_x\times d_x^2$ coefficients matrix built from columnwise Hadamard products.  Both of these are expanded upon further in Eq.~(\ref{genwout})-(\ref{polyvect}), and the form of $a_{2,(i,j)}$, Eq.~(\ref{prod3}) is derived in Eq.~(\ref{prod2}). 

We summarize in this brief section that the discussion of nonlinear quadratic VAR from a linear RC with Hadamard-quadratic read-out is similar to that of linear VAR with linear RC and linear read-out as discussed in Sec.~\ref{ReviewLinRC}. However, Eq.~(\ref{ls4}) generalizes, so that we may still write, 
$\bW^{out}:=\bv^*=\ba^* {\mathbb A}_\lambda^\dagger=\bY {\mathbb X}_\lambda^\dagger {\mathbb A}^\dagger_\lambda,$ but now ${\mathbb A}=\begin{bmatrix} {\mathbb A}_1\\{\mathbb A}_2\end{bmatrix}$.  In this statement, ${\mathbb A}_1$ is a renaming of what was ${\mathbb A}$ in Eq.~(\ref{Akry}), but now ${\mathbb A}_2$ is the $dr\times k d_x^2$ matrix defined in Eq.~(\ref{A2}) in the Appendix, and likewise ${\mathbb X}=\begin{bmatrix} {\mathbb X}_1\\{\mathbb X}_2\end{bmatrix}$ defined in Eq.(\ref{X2}).

While the linear reservoir + linear readout = VAR forecasting seems to only give reasonable results for short time forecasting, Fig.~\ref{fig:rcl2}, linear reservoir + quadratic readout = NVAR seems to give better and longer range forecasting which also seem to remain true to the statistic of the chaotic attractor once errors have swamped the point forecasts.  See Fig.~\ref{fig:rcl3}

}

\section{Is There a Connection to DMD-Koopman?}\label{IsKoop}

To briefly answer the question titling this section, the answer is yes, there is a connection between VAR and DMD, and so to RC. The more nuanced answer is that the connection is not complete.  
Throughout the discussion so far, a specialized version of an RC using an identity activation function, yields a linear process that is shown to relate to a VAR that is also a linear process.  In this section we ask if it also relates to the dynamic mode decomposition, DMD \cite{schmid2010dynamic,rowley2009spectral, kutz2016dynamic,williams2015data}, a 
concept that is also premised on a linear process model  as a finite estimation of the infinite dimensional linear action of the Koopman operator on a function space of observables \cite{arbabi2017ergodic}.  In Koopman theory, instead of describing the evolution and geometry of orbits in the phase space, the transfer operator methods generally describe evolution of functions whose domain is the phase space \cite{lasota2013chaos, bollt2013applied}.  Recently this approach has excited a huge trend in applied dynamical systems, with many excellent research papers \cite{williams2015data,kutz2016dynamic,li2017extended,bollt2019geometric}, review papers \cite{arbabi2017ergodic,budivsic2012applied}, and  books \cite{kutz2016dynamic} toward theory, numerical implementation and scientific application practice.  Our focus here will remain narrow, the goal being to simply identify a connection to the RC and its related VAR, as discussed above.  A primary purpose of DMD methods are for modal analysis of the system to describe coherent and typical behaviors, but it also can be used for forecasting, and for this sake the analogy is drawn here.

For direct comparison, first allow some minor manipulations to relate  the VAR(k), Eq.~(\ref{genvarp}) and  Eq.~(\ref{fit}), to a typical DMD form.
A time-delay version of a linear evolution is a special case of an exact DMD written as follows, with notation used as above,
\begin{equation}\label{fit2}
\begin{bmatrix}
 |& | & \vdots &  |\\ 
\bx_{k+1}& \bx_{k+2} & \ldots & \bx_{N}  \\ 
|& | & \vdots &  |\\ 
 \bx_{k}& \bx_{k+1} & \ldots &  \bx_{N-1} \\ 
|& | &\vdots & |\\ 
\vdots & \vdots & \vdots & \vdots \\
|& | &\vdots & |\\ 
\bx_{2} & \bx_{3} & \ldots &  \bx_{N-k}\\
|& | & \vdots &  |\\ 
\end{bmatrix}
=
{\cal K}
\begin{bmatrix}
 |& | & \vdots &  |\\ 
\bx_k& \bx_{k+1} & \ldots & \bx_{N-1}  \\ 
|& | & \vdots &  |\\ 
 \bx_{k-1}& \bx_k & \ldots &  \bx_{N-2} \\ 
|& | &\vdots & |\\ 
\vdots & \vdots & \vdots & \vdots \\
|& | &\vdots & |\\ 
\bx_{1} & \bx_{2} & \ldots &  \bx_{N-k-1}\\
|& | & \vdots &  |\\ 
\end{bmatrix},
\end{equation}
or simply,
\begin{equation}\label{dmdmatrix}
{\mathbb X}'={\cal K} {\mathbb X},
\end{equation}
where 
$ {\mathbb X}$, and ${\mathbb X}'$ are the $k d_x \times (N-k-1)$ data matrices in Eq.~(\ref{fit5}), and ${\cal K}$ is a $k d_x\times k d_x$ DMD matrix approximating the action of the infinite-dimensional Koopman operator.  
Abusing notation slightly, the least squares problem, 
\begin{equation}
{\cal K}=\argmin_K \|{\mathbb X}'-K {\mathbb X}\|_F,
\end{equation}
 has the solution,
\begin{equation}\label{exact}
{\cal K}= {\mathbb X}' {\mathbb X}^\dagger ,
\end{equation}
which is called the ``exact DMD" solution. 
While there are many variants of DMD, 
this one called exact DMD is popular for its simplicity of implementation while still useful for interpreting the system in terms of modal behaviors.


Contrasting ${\cal K}$ derived by  exact DMD, Eq.~(\ref{fit2}), versus ${\cal A}$ for the VAR(1) form described in Eqs.~(\ref{var1form})-(\ref{fit5}) reveals clear similarities since each states a linear relationship between the same data,
${\mathbb X}'={\cal K} {\mathbb X}$, versus ${\mathbb X}'={\cal A}{\mathbb X}$, but these are ill-posed equations and the ${\cal A}$ need not be the same as ${\cal K}$.  Closer inspection reveals that  Eq.~(\ref{exact}) allows freedom for best least squares fit considering the entire matrix ${\cal K}$, and so differences relative to  
Eqs.~(\ref{fit}),  (\ref{ls2}).  Whereas, only the first $k$ rows of ${\cal A}$ are free parameters in the regression; the subsequent rows of ${\cal A}$ are sparsely patterned with either zero's or the identity matrix, Eq.~(\ref{var1form}).

A similar, but not identical, structural difference appears when contrasting the SVD based exact DMD to the original DMD method of Schmidt \cite{schmid2010dynamic} and also Rowley and Mezic, \cite{rowley2009spectral} which is an Arnoldi-like version of DMD in terms of iterations in a Krylov space \cite{arnoldi1951principle, van2003iterative}. Reviewing that Arnoldi-version of DMD, using the notation of \cite{rowley2009spectral}, observations $\bx_k\in {\mathbb R^d}$ are assumed (fitted) to be from a linear process, 
but also by considering the iterations are to be fitted in the Krylov space, assuming that $\bx_m\in Kry_m(\bx_0)=span\{\bx_0,A\bx_0,...,A^{m-1}\bx_0\}$ for data, $K=[\bx_0 |\bx_1| \ldots |\bx_m]=[\bx_0 | A \bx_0 | \ldots |A^{m-1} \bx_0]$.  Stating the linear combination $\bx_m=A\bx_{m-1}=c_0 \bx_0+ \ldots +c_{m-1}\bx_{m-1}=K\bc$, where $\bc=[c_0;c_1;\vdots;c_{m-1}]$ is the vector of coefficients.  Then a key and clever observation was to rewrite this in terms of a companion matrix, 
\begin{equation}\label{companion}
C=
\begin{bmatrix}
0 & 0 & \ldots & 0 & c_1 \\
1 & 0 &   & 0 & c_1 \\
0 & 1 & & 0 & c_2 \\
\vdots & \ddots & & \vdots \\
0 & 0 & \ldots & 1 & c_{m-1}
\end{bmatrix}.
\end{equation}
So that results,
\begin{equation}
AK=KC.
\end{equation}
From there, exploiting the theme of Arnoldi methods, the eigenvalues of $C$ are related as a subset of the eigenvalues of $A$ and with a direct linear relationship between eigenvectors of $C$ and $A$, Ritz vectors, and the unknown coefficients $\bc$ of $C$ can be computed by a least squares procedure.  Keeping in mind that power iterations as one does in Krylov spaces emphasize just the dominant direction, the Arnoldi methods take care to orthogonalize at each step in the algorithm for stabilization an otherwise unstable search for large sparse matrices, and these make deliberate use of QR decompositions.  
Our interest here is only to point out analogies between ${\cal A}$ from reservoir computing and VAR(1) and ${\cal K}$, rather than to continue toward discussion of modal analysis as one does in DMD analysis.    Summarizing, the analogy we see that the companion matrix $C$ in Eq.~(\ref{companion}) reminds us of the companion matrix ${\cal A}$ in Eq.~(\ref{var1form}).  However the most significant difference is that while $c_i$ are scalars, that $\ba_i$ are 
$d_x\times d_x$ matrices.

 \section{Examples}\label{examples}
 
 The example Figs.~\ref{fig:LSM1}, \ref{fig:LSM2}-\ref{fig:LSM4} already threaded in the above presentation of methods, were in terms of the Mackey-Glass differential delay equation system, which we now recall.  Then in the subsequent, we will show similar figures highlighting the concepts in a different system, the famous Lorenz63 ODEs.

 \subsection{Example 1: Mackey-Glass Differential Delay Equation}\label{MGEsec}

 The Mackey-Glass differential delay equation, \cite{mackey1977oscillation}, 
\begin{equation}\label{MGE}
x'(t)=\frac{a x(t-t_d)}{1+[x(t-t_d)]^c}-b x(t),
\end{equation}
has become a now classic standard example in time-series analysis \cite{farmer1982chaotic,lichtenberg85j} of a high (infinite) dimensional dynamical system with a low-dimensional attractor, which we have used as a benchmark in our own previous work for machine learning \cite{bollt2000model} dynamical systems.    
The problem is physiologically relevant for describing dynamic diseases.
A differential delay equations can be described as infinite dimensional dynamical systems, a concept that is more easily understandable in terms of the notion that an initial condition state advances not just a finite dimensional vector, but rather an entire interval $[t_0,t_0+t_d]$ of initial values of $x(t)$ are required.  However, the MG equations have a nice property for the its practical use as a benchmark problem, which is  that there is essentially an attractor whose fractal dimension varies with respect to the parameters chosen, allowing for a complexity tunable test problem.
We have chosen parameters $t_d=17, a=0.2, b=0.1, c=10.0$ for which if pursing time delay embedding gives an embedding dimension of $d=4$.   We use integration steps of $\Delta t=t_d/100$ throughout.  
We show  time-series in Fig.~\ref{fig:LSM1}, a standard nonlinear RC forecast of the system in Fig.~\ref{fig:LSM2}, and the linear RC/VAR forecast of the system in Figs.~\ref{fig:LSM3}-\ref{fig:LSM4}.


\begin{figure}[htbp]
\centering
\includegraphics[scale=.65]{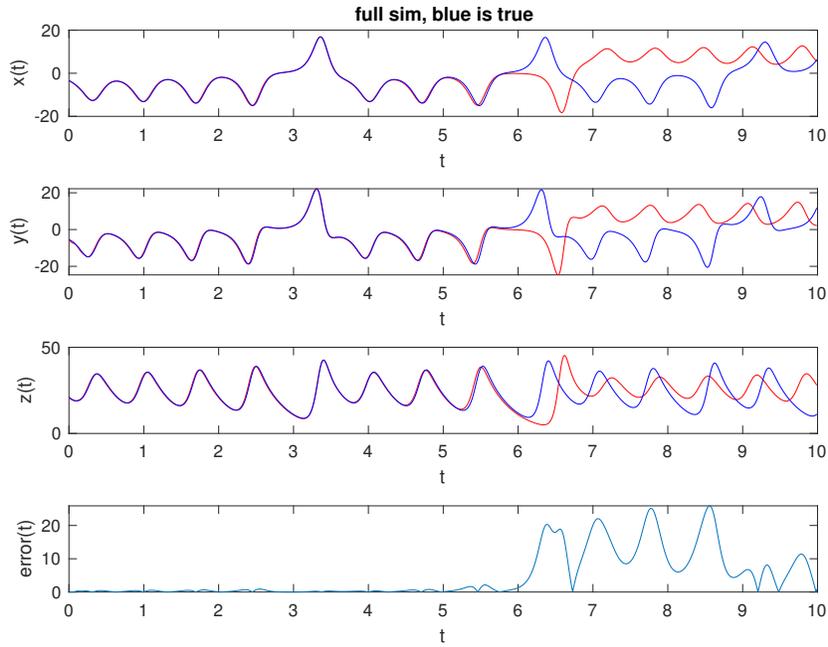}
\includegraphics[scale=.62]{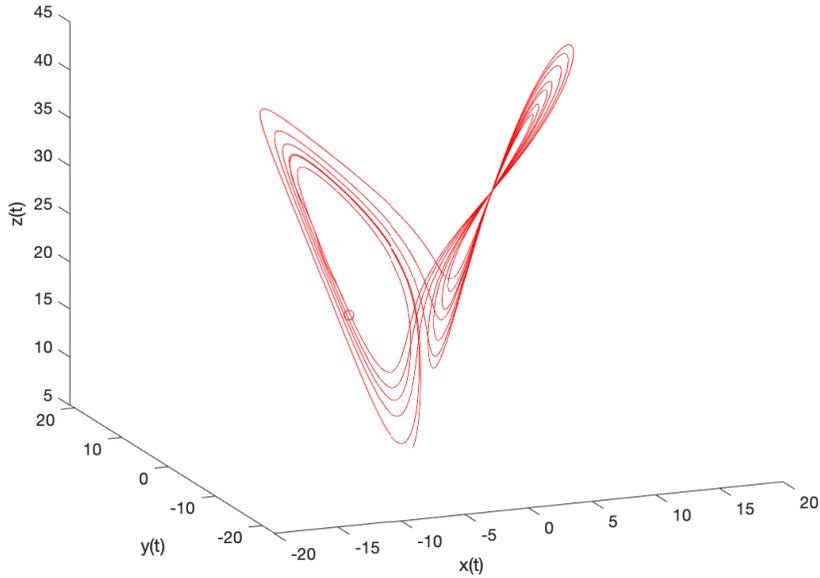}
\caption{ {\color{black} Lorenz time-series with nonlinear RC, so with $q(x)=tanh(x)$, of size $d_r=1000$,  and quadratic read-out, $\bW^{out}$ fitted to input $\bR=[\bR_1; \bR_2]$, 
 from Eq.~(\ref{newR}). }  The top three time-series are state variables and blue curves show forecasts where red shows true data.  Bottom, error shown, growing from initial seed. (Right) The phase space presentation of the forecast variables, (x(t),y(t),z(t)).  Error performance is excellent.  And, even once error has grown,  the produced attractor seems true to the forecast Lorenz. Compare to Figs.~\ref{fig:rcl2}-\ref{fig:rcl3}. }
\label{fig:rcl1}
\end{figure}

\begin{figure}[htbp]
\centering
\includegraphics[scale=.65]{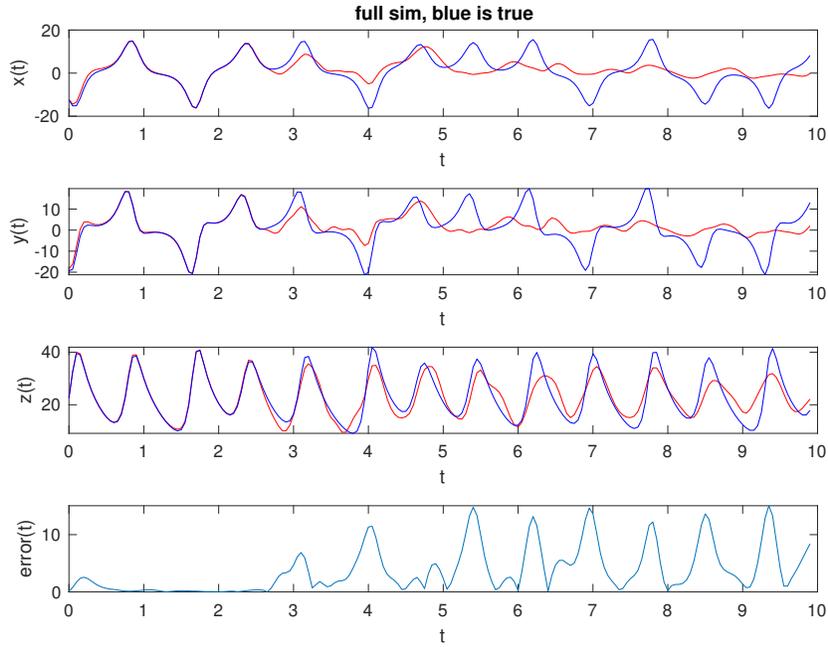}
\includegraphics[scale=.62]{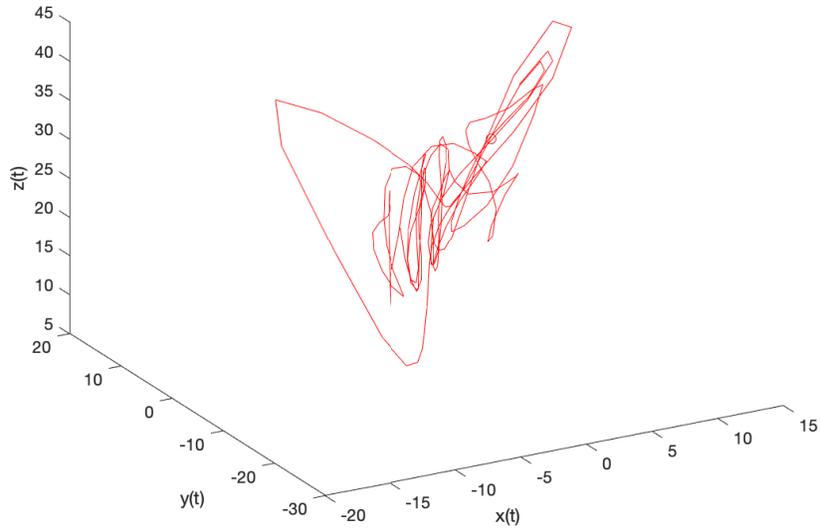}
\caption{{\color{black} Lorenz time-series with fully linear RC, so $q(x)=x$ of size $d_r=1000$, and linear read-out. } (Left) the top three time-series are state variables and blue curves show forecasts where red shows true data.  Bottom, error shown, growing from initial seed.  This fully linear RC, being equivalent to a VAR does produce good forecasts for a finite time, even if for a shorter time than in the nonlinear methods of Figs.~\ref{fig:rcl1}, \ref{fig:rcl3}.
(Right) The phase space presentation of the forecast variables, (x(t),y(t),z(t)).  Now unlike the nonlinear RC or the nonlinear read-out cases, despite forecasts for a little while, once errors have occurred in forecasting the time-series of the fully linear RC, the form of the attractor is entirely wrong. }
\label{fig:rcl2}
\end{figure}

\begin{figure}[htbp]
\centering
\includegraphics[scale=.65]{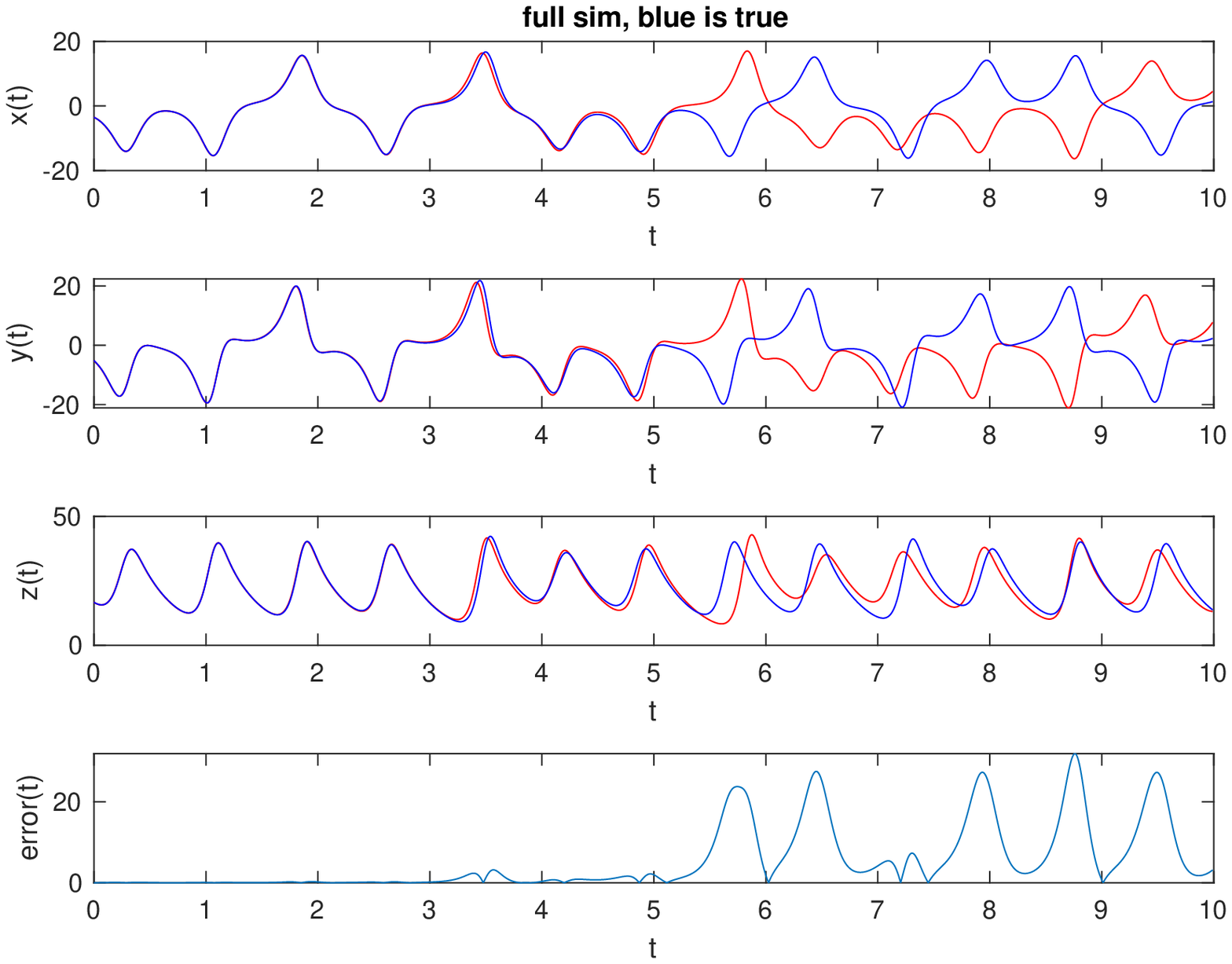}
\includegraphics[scale=.62]{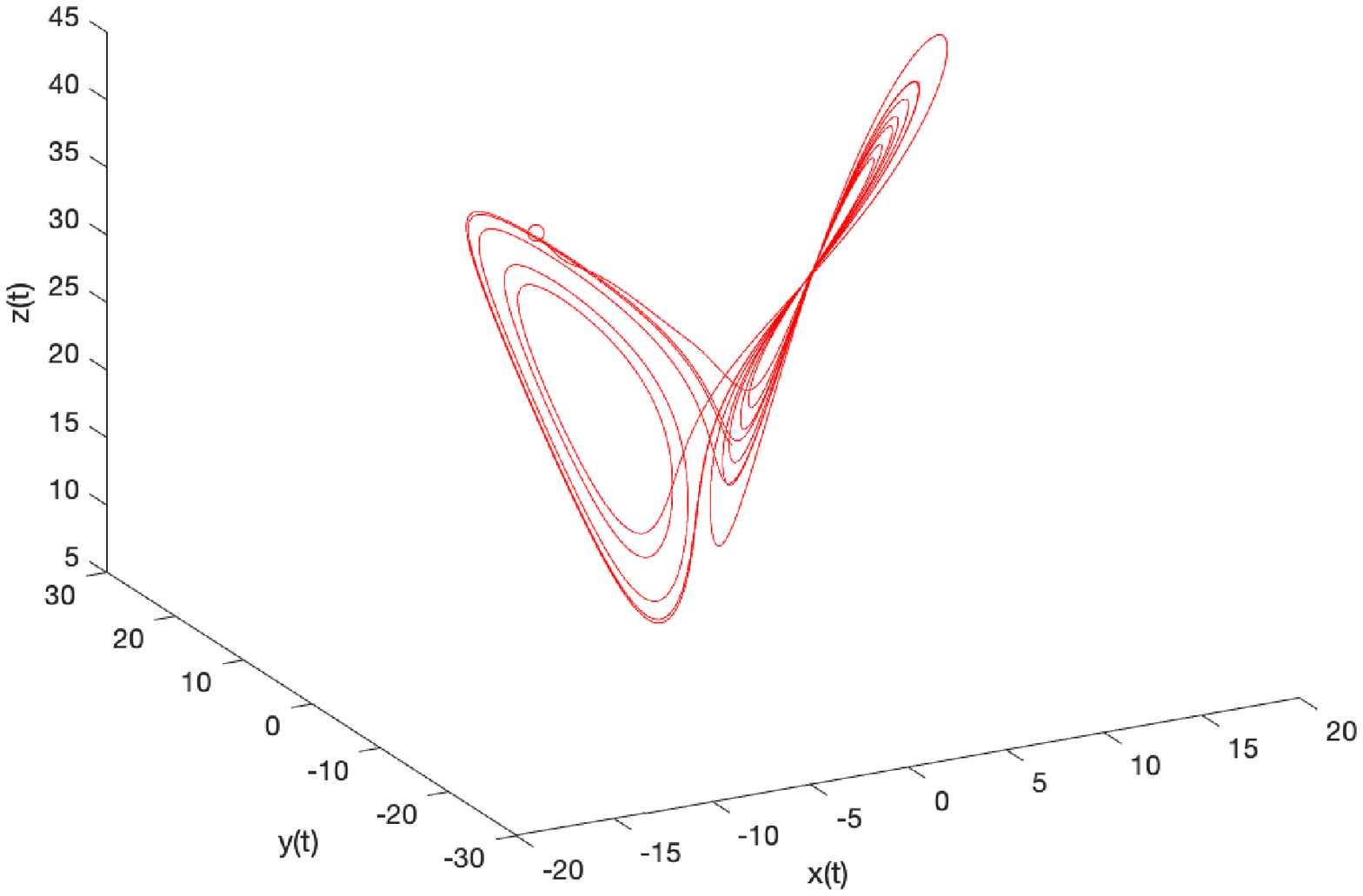}
\caption{{ \color{black} Lorenz time-series with linear RC, so $q(x)=x$ of size $d_r=1000$, but quadratic read-out, $\bW^{out}$ fitted to input $\bR=[\bR_1; \bR_2]$, 
 from Eq.~(\ref{newR}). } (Left) the top three time-series are state variables and blue curves show forecasts where red shows true data.  Bottom, error shown, growing from initial seed.  (Right) The phase space presentation of the forecast variables, (x(t),y(t),z(t)).  As in full nonlinear RC-method of Fig.~\ref{fig:rcl1}, the attractor seems true even once errors have grown. Compare to Figs.~\ref{fig:rcl1}-\ref{fig:rcl2}.}
\label{fig:rcl3}
\end{figure}

 \subsection{Example 2: Lorenz63}\label{lorensec}
 
The Lorenz63 system \cite{lorenz1963deterministic} is the three coupled ordinary differential equations:
\begin{eqnarray} \label{eq:lorenz}
\dot{x} & = & 10 (y-x), \nonumber \\
\dot{y} & = & x(28-z)-y, \nonumber \\
\dot{z} & = & xy-(8/3)z. 
\end{eqnarray}
While these Lorenz equations may have been originally posed as time varying Fourier coefficients for describing a partial differential equation system describing  convection rolls of heated fluid in an atmospheric system, they have become  been a popular paradigm in the study of chaotic systems, for foundation principles of chaos historically and ongoing, as a simple and familiar benchmark problem and also in the pedagogy of dynamical systems.   
The chaotic attractor in the phase space 
$\{x(t),y(t),z(t)\}$ illustrates a familiar butterfly, but we show a segment of the $x(t)$ time series that will be used as our data set, in Figs.~\ref{fig:rcl1}-\ref{fig:rcl3} .  Also shown are nonlinear RC, $q(x)=tanh(x)$ activation forecasts in Fig.~\ref{fig:rcl1} using the usual nonlinear reservoir computing, with excellent success.  In Fig.~\ref{fig:rcl2}, we show forecasting results using a linear $q(x)=x$ activation RC with still good results for short term forecasting agreeable with the expectation with  VAR theory.

{\color{black} Consider the following summary of results of three experiments shown in Figs.~\ref{fig:rcl1}-\ref{fig:rcl3}.
\begin{enumerate}
\item Fig.~\ref{fig:rcl1} shows a standard fully nonlinear RC, with activation $q(x)=tanh(x)$ and Hadamard quadratic read-out  $\bW^{out}$ of the reservoir variable $[\br; \br\circ \br]$.  Forecasting is excellent, and interestingly, apparently even once errors have accumulated, the RC continues to produce a Lorenz-like attractor.
\item Fig.~\ref{fig:rcl2} shows a fully linear RC with linear activation, $q(x)=x$, and linear readout $\bW^{out}$  of the reservoir variable $\br$ is equivalent to a VAR and it produces good short term forecasts, but for  shorter time than the fully nonlinear RC.  Also once errors accumulate, the long term forecasts do not produce a Lorenz like attractor, but rather seem to converge to zero.
\item Figs.~\ref{fig:rcl3} shows a linear RC, so $q(x)=x$, but with a Hadamard quadratic read-out  $\bW^{out}$ of $[\br; \br\circ \br]$.  Forecast performance is good and the attractor seems to be well reproduced even once error has grown, comparably to the fully nonlinear case of Fig.~\ref{fig:rcl1}.
\end{enumerate}
}

{\color{black}
\section{On Fading Memory}\label{vanishing}

There are several ways to consider memory in the system.    Memory as it turns out is an important property for an echo state machine since it was recently shown by Gonan and Ortega, \cite{gonon2020fading} that these are universal.  The VAR representation shown here allow for own discussion of this property. 

\subsection{Fading Memory In Terms of the Role of the Internal Layers} 
One way is to think of the connection between the reservoir and the coefficient matrices $\ba_j$ may become increasingly small in a way that is moderated in part by the randomly chosen $\bA$, yields a fading bound on the true memory.  From Eq.~(\ref{prod1}), $a_{j}=\bW^{out} \bA^{j-1}\bW^{in}$, follows a bound, in norm, 
\begin{eqnarray}\label{prodd2}
    \|a_{j}\|_\star & = &\|\bW^{out} \bA^{j-1}\bW^{in} \|_\star \nonumber \\
   &\leq &  \|\bW^{out}\|_\star \| \bA^{j-1} \|_\star \| \bW^{in} \|_\star \nonumber \\
   &\leq &  \|\bW^{out}\|_\star \| \bA \|_\star^{j-1} \| \bW^{in} \|_\star.
\end{eqnarray}
$\| \cdot \|_\star$ denotes an induced matrix norm  inherited by the corresponding  vector norm.  That is, allowing a (possibly not square) matrix $B$ mapping between vector spaces $V_1, V_2$, by  $B:V_1\rightarrow V_2$, then $\|B\|_{\star, V_1,V_2}:=\sup_{\|x\|_{\star,V_1}=1} \|B x\|_{\star,V_2}$, and $\|\cdot \|_{*,V_i}$ describes the vector norm in $V_i$.  For example, typical  favorite vector norms include the Euclidean norm, $\|\cdot \|_2$, the 1-norm, $\|\cdot \|_1$ and the $\infty$-norm, $\|\cdot \|_\infty$.  The specific  vector norm is not important,  thus the noncommital notation $\star$, for good reason to be described in a moment.  For simplicity of notation in Eq.~(\ref{prod2}), we omit emphasis of  domain $V_1$ and range $V_2$ vector spaces of each matrix operator, understanding that dimensionality of the induced vector norms depends on the matrix sizes and ranks.  

Inspecting the term, $ \| \bA \|_\star^{j-1}$, in Eq.~(\ref{prodd2}), we can bound by its eigenvalues.  For any chosen $\epsilon>0$, \cite{golub2013matrix}, there exists some induced matrix norm, $\| \cdot \|_\star$ (which is why we used the noncommittal notation $\star$) such that $ \| \bA \|_\star \leq \rho(\bA)+\epsilon$, where $\rho(\bA)=\max_{\{\lambda_i\}} |\lambda_i|$ is the spectral radius defined as the largest magnitude of eigenvalues, noting that $\bA$ is square.  By  theorem \cite{golub2013matrix},  if $\rho(\bA ) <1$ then there exists an induced norm such that in terms of that norm, $\|\bA\|_\star <1$, which implies $\|\bA^n \|_\star \leq \|\bA \|_\star^n \rightarrow 0$ as $n\rightarrow \infty$.  Therefore, by equivalence of norms, $\bA^n \rightarrow 0$.  This last statement is stronger than a convergence in norm, but rather it is is a componentwise statement of convergence of the matrix to the zero matrix with respect to exponentiation, under the condition of spectral radius bounded by 1.

A result  regarding the VAR matrix mentioned in Eqs.~(\ref{prod1}), (\ref{prodd2}), that as long as the read-out matrix is bounded $\|\bW^{out} \|<C <\infty$, and $\rho(A)<1$, then coefficient matrices become increasingly close to the zero matrix, 
\begin{equation}
a_j \rightarrow 0, \mbox{ as }j\rightarrow \infty,
\end{equation}
stated  strongly as componentwise convergence  of the matrices.  However, this is not to say that convergence need to be monotone, or that the bound must sharp.
What is interesting in this statement is that  the distribution from which the randomly chosen  matrix $\bA$ is drawn controls the spectral radius $\rho(\bA)$, which in turn dominates the VAR matrices $a_j$.  This is agreeable with the general VAR theory reviewed in Sec.~\ref{generalvar}, that a general VAR($\infty$) is well approximated by a finite VAR(k).



\subsection{Experiments on Fading Memory of the RC} 

Consider the actual computed $a_j$ matrices, and their corresponding induced norms $\|a_j\|_2$ as computed from the example of chaotic systems studied in the previous section, which were the Mackey-Glass system, and the Lorenz system.

In Fig.~\ref{fig:vm}(Top), there is shown a ($k$, $\|a_k\|_2$) curve as computed from Eq.~(\ref{prodd2}) in the case of a RC model derived from the Mackey-Glass equation, corresponding to RC forecasts of the same system in Fig.~\ref{fig:LSM1}.  We see a pronounced fading memory as the $\|a_k\|_2$ have diminished to negligable values by $k\geq 6$.  This is not unexpected since the attractor of these equations shown in Fig.~\ref{fig:LSM1}, illustrates primarily rotation and otherwise weak chaos, in the sense that while the embedding dimension may be $d=4$, the Lyapunov exponent is relatively small even if positive, $\lambda_{max}=0.0058$, \cite{wernecke2019chaos}.

Considering the memory curve ($k$, $\|a_k\|_2$) of the Lorenz system in Fig.~\ref{fig:vm}(Bottom), we see that initially $\|a_k\|_2$ decreases quickly, to small values by $k=9$ but then the computed values fail to decrease further for larger $k$.  We attribute this to difficulties of long range forecasting of this highly chaotic system.  Another way of realizing this point is that the details of the tail, of the large $k$ $a_k$ values vary with different data samples even though the front part of the series (small k) are stable across samples.  Interestingly, nonetheless, as an affirmation of the theoretical discussion, the green curve derived from q.~(\ref{prodd2}) involving the RC, and the blue curve derived as a VAR fit, Eq.~(\ref{ls2}), almost entirely coincide, the difference being likely due to numerical estimation issues for very large $k$.

\begin{figure}[htbp]
\centering
\includegraphics[scale=.5]{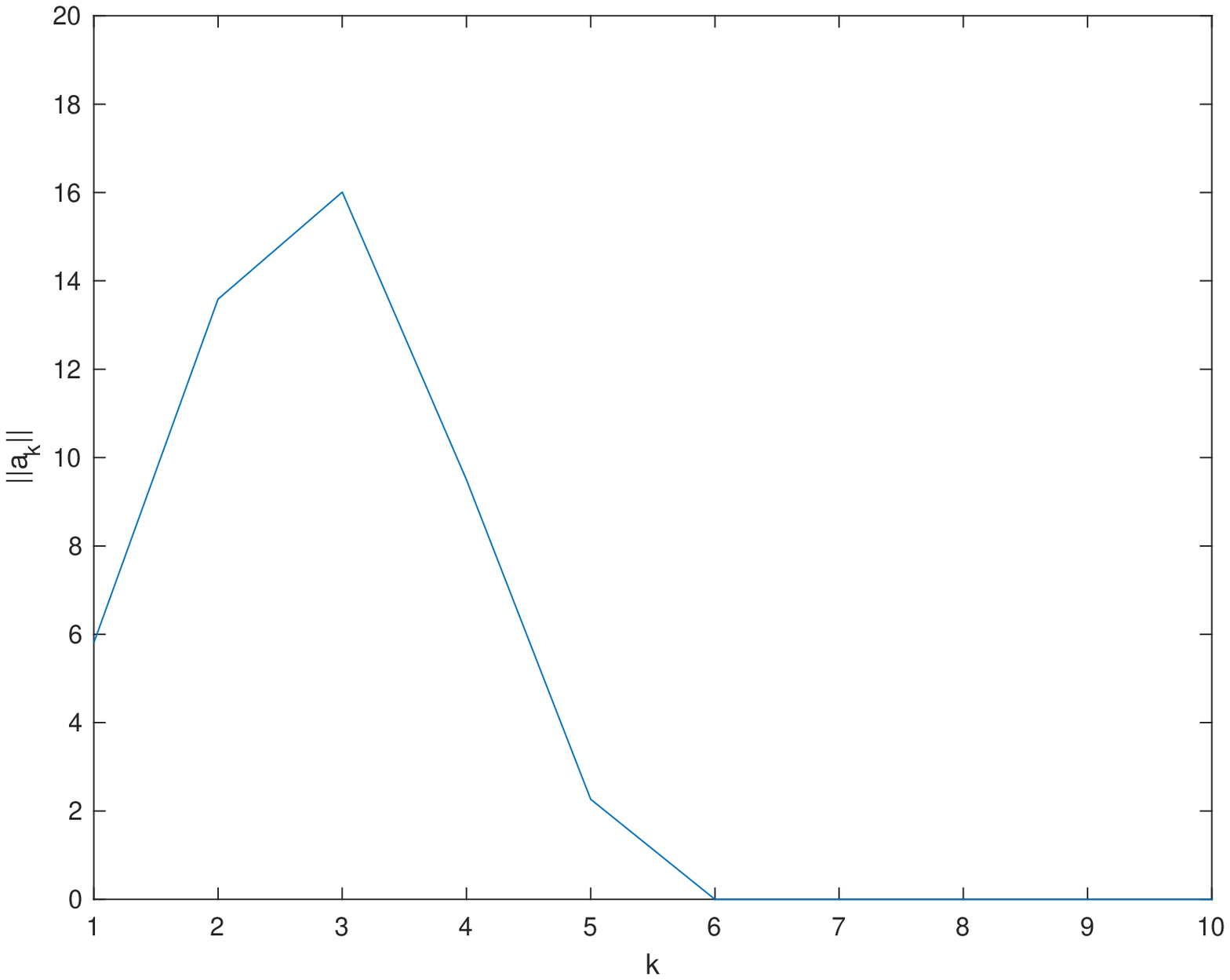}
\includegraphics[scale=.32]{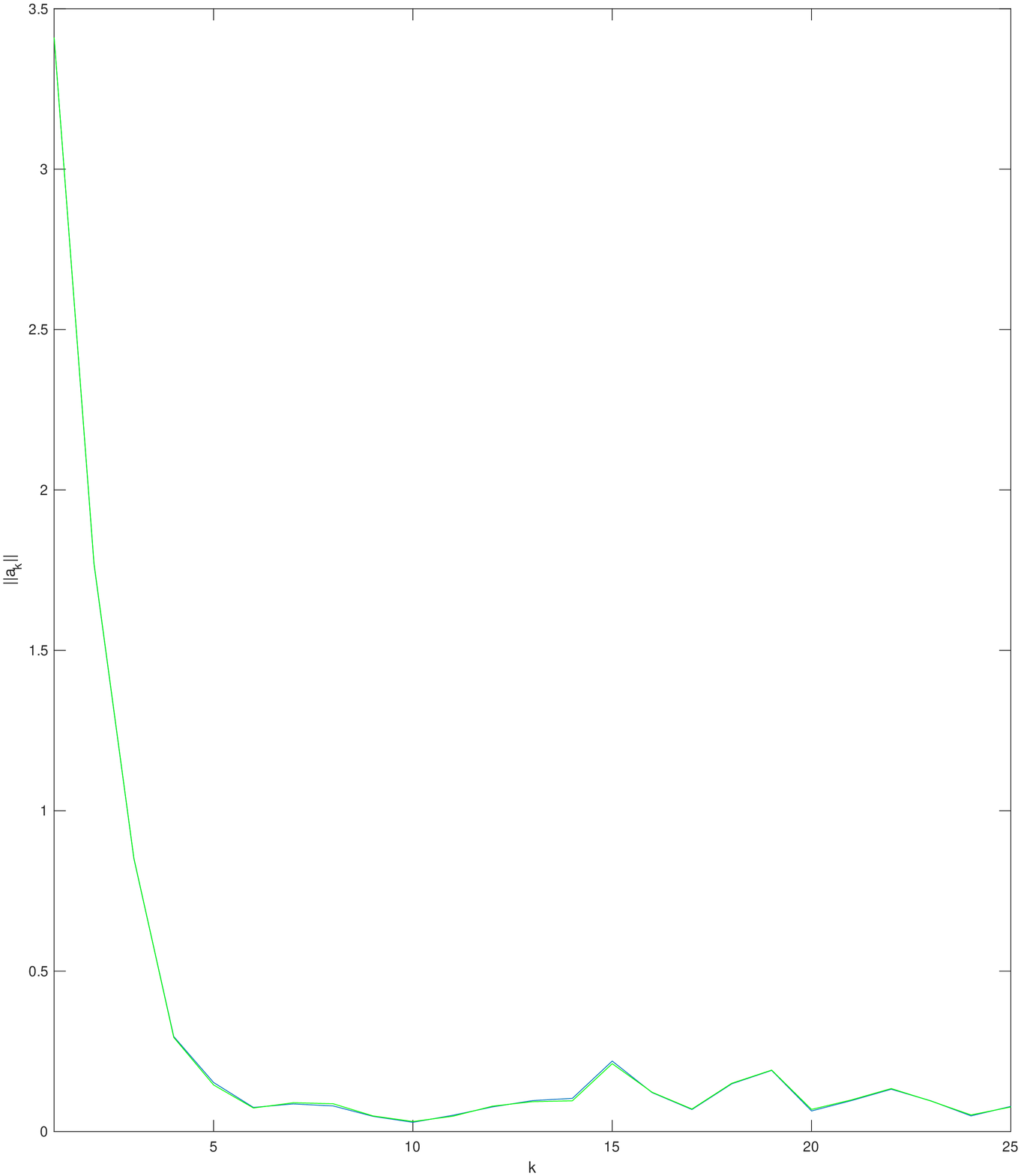}
\caption{{\color{black} Fading memory as observed by magnitude of VAR(k) matrices $a_k$, as $k$ vs $\|a_k\|_2$ induced norm, of (Top) Mackey-Glass system, the VAR(6) model is  descriptive since $\|a_6\|_2$ is already relatively small. (Bottom) Lorenz system where we see that after $k=9$ coefficient matrices $\|a_9\|_2$ are relatively small, suggesting the VAR(9) description is close to the large $k$ model.   However, we do not see convergence to zero suggested by Eq.~(\ref{prodd2}), which we attribute to numerical instabilities since the details of the computed $a_j$ seem to vary with details of the orbit data, whereas the head (the early terms) remain stable.  Shown in green curve are the $a_k$ computed by Eq.~(\ref{prodd2}) involving the RC, but  the blue curve directly as a VAR fit, Eq.~(\ref{ls2}), and we see that these closely agree as the two curves are almost coincident.  } }
\label{fig:vm}
\end{figure}
}

\section{Conclusion}

The success of machine learning and artificial neural networks has lead to a clear and overwhelming widely adopted wave across so many areas where data is relevant and patterns are of interest.  Dynamical systems is no exception and forecasting a dynamical system is a specific application that is broadly relevant and of interest to us here.  The RNN framework is particularly relevant for a dynamical systems since the reserve memory aspect of the concept allows for a good framework and some aspects of delay embedding.  However while the RNN tends to have many many parameters to fit, with the danger of overfitting always present, and in any case the large cost of optimization in the training phase, there is a surprising short cut.  The echo-state/reservoir computing concepts presume to choose the weights for input layer and internal layer entirely randomly.  Then only the output layer is trained.  Furthermore that output layer training will be by linear algebraic manipulations toward a least squares solution, rather than the usual nonlinear optimization necessary for the many parameters of the full nonlinear RNN.  That this would allow a huge computational savings is clear.  What is perhaps  a surprise is how this gross simplification still yields useful results.  While there have been a number of studies  experimentally describing how to choose better random processes to define the random parameters, e.g. such as to emphasize sparsity, or to control the spectral radius, and other properties, in this work we have taken a different approach, which is not specifically to improve the performance of the concept but instead to give a partial explanation as to how the concept can work at all.  After all, at first glance it may seem that it would be impossible that such a simplification could work.  In this work, we have simplified the RC concept, allowing for the activation function to be an identity function instead of the more typical sigmoidal function.  In this case, the process is entirely linear and so easier to study.  As it turns out the RC still performs reasonably well {\color{black} for short term forecasts},    and it is certainly leads to easier to analysis.  Herein we  prove that the linear RC is in factly directly related to the more matured topic of VAR, vector autoregressive time series forecasting, and with all the related theory including  the WOLD theorem as a representation theorem, which therefore now applies to the RC.  Also we are able to make a direct connection to the increasingly popular DMD theory.  
{\color{black} Further, the  commonly used fitting upon readout of linear and Hadamard quadratic observations of the reservoir states yields a nonlinear VAR (NVAR) allowing for all quadratic monomial terms generalization of the VAR result.  This NVAR version apparently not only makes competitive forecasts, but also errors seem to respect the original attractor in the case of the Lorenz system.}


\section{Data Availability Statement}

The data that support the findings of this study are available from the corresponding author upon reasonable request.

\section{Acknowledgments}

The author received funding from the Army Research Office (N68164-EG) and also DARPA.  I would also like to sincerely thank Daniel J. Gauthier, Aaron Griffith, and Wendson A. S. Barbosa of Ohio State University for their generous and helpful feedback concerning this work.

\section{Appendix: Review of Regularized Pseudo-Inverse} \label{appmtrix}

We review how to numerically and stably 
compute the pseudo-inverse by the singular value decomposition, with  regularized singular values (SVD).  
Reviewing the matrix theory of regularized pseudo-inverses for general matrices,  if,
\begin{equation}
Xb=z,
\end{equation}
 $X_{n\times p}, b_{p\times 1}, z_{n\times 1}$, then if the SVD is $X=U \Sigma V$, with orthogonal matrices, $n\times n$, $U$ satisfies, $U U^T=U^T U=I$ and $p\times p$, $V$ satisfies  $V V^T=V^T V=I$, and $\Sigma$ is $n\times p$ ``diagonal" matrix of singular values, $\sigma_1\geq \sigma_2\geq \sigma_r \geq 0 \geq \sigma_p\geq 0$,
\begin{equation}
\Sigma=
\begin{bmatrix}
\sigma_1 & & \\
& \ddots & \\
& & \sigma_p
\end{bmatrix}, \mbox{ if }n=p,
\Sigma=
\begin{bmatrix}
\sigma_1 & & \\
& \ddots & & \vdots \\
& & \sigma_p
  &  & 
\end{bmatrix}, \mbox{ if }n >p,
\begin{bmatrix}
\sigma_1 & & & \ldots \\
& \ddots & & \ldots \\
& & \sigma_p & \ldots
\end{bmatrix}, \mbox{ if }n<p.
\end{equation}
Then,
\begin{equation}
X^\dagger:=(X^T X)^{-1}X^T=V \Sigma^\dagger U^T,
\mbox{ where, }
\begin{bmatrix}
\frac{1}{\sigma_1} & & \\
& \ddots &   \\
& & \frac{1}{\sigma_p} \\
\vdots & \vdots & \vdots
\end{bmatrix}, \mbox{ in the }n>p\mbox{ case }.
\end{equation}
The least squares estimator of $Xb=z$ is,
\begin{equation}
b^*=(X^TX)^{-1}X^T z:=X^\dagger z,
\end{equation}
and we write the ridge regression Tikhonov regularized solution,
\begin{equation}\label{tik1}
b_\lambda^*=(X^TX+\lambda I)^{-1} X^t z=V(\Sigma^T\Sigma +\lambda I)^{-1} \Sigma^T U^T z:=X_\lambda^\dagger z.
\end{equation}
The regularized pseudo-inverse $X_\lambda^\dagger$ is better stated in terms of the regularized singular values, by,
\begin{equation}\label{tik2}
\Sigma_\lambda^\dagger:=(\Sigma^T\Sigma +\lambda I)^{-1} \Sigma^T=
\begin{bmatrix}
\frac{\sigma_1}{\sigma_1^2+\lambda} & & \\
& \ddots & \\
& & \frac{\sigma_p}{\sigma_p^2+\lambda} \\
\vdots & \cdots & 
\end{bmatrix} \mbox{ in the } n> p \mbox{ case,}
\end{equation}
and then,
\begin{equation}
b^*_\lambda=X^\dagger_\lambda z=V\Sigma_\lambda^\dagger U^T z.
\end{equation}
Throughout, since we will always refer to regularized pseudo-inverses, we will not emphasize this by abusing notation allowing that $b^*$ denotes $b^*_\lambda$ even if only a very small $\lambda>0$ is chosen, unless otherwise stated, $\lambda=1.0 \times 10^{-8}$.  This mitigates the tendency of overfitting or likewise stated in terms of zero or almost zero singular values that would otherwise appear in the denominators of $\Sigma^\dagger$.  The theory is similar for $n=p$ and $n<p$, as well as the scenario where $z$ is not just a vector but a matrix, and likewise as in Eq.~(\ref{RR}) where we refer to the transpose scenario.

{\color{black}
\section{Appendix: On Quadratic NVAR Connection to RC}\label{nvarapp}

In this appendix, we give details claimed in Sec.~\ref{NLinRC} that a linear RC with the Hadamard quadratic nonlinear read-out, also corresponds to a VAR like entity from stochastic process modeling of time-series.  Now, however a quadratic type nonlinear VAR results, an NVAR.  This is a generalization of the linear VAR discussion of Sec.~\ref{ReviewLinRC}

A commonly used scenario of RC \cite{lu2018attractor} is to fit $W^{out}$ not just to $\br$ data, but also to $\br\circ \br$, where $\circ$ denotes the Hadamard product (implemented in array languages such as Matlab by ``array arithmetic" using the `$.^*$' notation with the dot, in place of what would otherwise be $`*'$ for standard matrix multiplication).  For a vector $\br=[r_1| r_2 |... |r_{d_r}]^T$,  this is defined as component-wise operations: $\br\circ \br=[r_1^2 |r_2^2|...|r_{d_r}^2]^T$.   The reason for using nonlinear terms is cited as improved performance, allowing for matching the parity of the process.   In this case, we rename what before we called $\bR$ to now be called $\bR_1$.  So,
Eq.~(\ref{Rdata}) is replaced with,
\begin{eqnarray}\label{newR}
\bR_1&=&
\left[
\begin{array}{cccc}
\br_k & | \br_{k+1} & | ... &| \br_N   
\end{array}\right],
\bR_2=\left[
\begin{array}{cccc}
\br_k \circ \br_k & | \br_{k+1} \circ  \br_{k+1} & | ...& |\br_N\circ \br_N 
\end{array}\right] \nonumber \\
\bR&=&
\left[
\begin{array}{c}
\bR_1 \\
\bR_2
\end{array}
\right].
\end{eqnarray}
Then
  Eq.~(\ref{RR}) remains written as before, $  \bW^{out}:=\bX \bR^T(\bR \bR^T+\lambda {\mathbf I})^{-1},
$ but now since $\bR$ is $2d_r\times N-k$, then $\bW^{out}$ is $d_x\times 2 d_r$.   For convenience of the rest of this section, partition these matrices $\bW^{out}$  into top and bottom half portions.  These, we show, act on linear and quadratic terms of the corresponding NVAR, 
\begin{equation}\label{genwout}
\bW^{out}=\begin{bmatrix} \bW_1^{out} \\ \bW_2^{out}\end{bmatrix},
\end{equation}
 each of size $d_x\times d_r$.
  
  First, note an identity of how the Hadamard product distributes with standard  matrix-vector multiplication.  Let $\bw=[w_1|w_2|...|w_n]^T$ a vector with scalar vector components $w_i$, and $B$ a general $m\times n$ matrix.  Let $B=[\bb_1 | \bb_2 | ... | \bb_n]$  written in terms of the column vectors $\bb_j$ of $B$.
  Then,
  \begin{eqnarray}\label{hadmat}
  B \bw \circ B \bw &=& (w_1 \bb_1+ w_2 \bb_2+...+w_n \bb_n) \circ (w_1 \bb_1+ w_2 \bb_2 +...+w_n \bb_n) \nonumber \\
  &=& [\bb_1\circ \bb_1| \bb_1 \circ \bb_2| ... | \bb_n \circ \bb_n] 
  \begin{bmatrix}
  w_1^2 \\ w_1 w_2 \\ \vdots \\w_1 w_n \\w_2 w_1 \\ w_2^2 \\ \vdots \\w_n^2
  \end{bmatrix}:=P_2(B,B)p_2(\bw,\bw).
\end{eqnarray}
Thus  the Hadamard operator distributes through matrix multiplication to be written purely as matrix multiplication with carefully stated matrices.
We have defined the matrix of Hadamard products as $P_2(B,B)$, which is a $m \times n^2$ matrix by the matrix function defined in Eq.~(\ref{hadmat}),
\begin{equation}
P_2:{\mathbb R}^{m \times n} \times{\mathbb R}^{m \times n}  \rightarrow {\mathbb R}^{m \times n^2},
\end{equation}
 and vector function also in Eq.~(\ref{hadmat}),
 \begin{eqnarray}\label{polyvect}
 p_2(\bv,\bw):{\mathbb R}^n\times {\mathbb R}^n &\rightarrow& {\mathbb R}^{n^2} \nonumber \\
 (\bv,\bw) &\mapsto& [v_1w_1| v_1w_2|...|v_1w_n|v_2w_1 |v_2w_2|...|v_nw_n]^T,
 \end{eqnarray}
  to be the $n^2\times 1$ vector of all quadratic combinations suggested in the equation above, $p_2:{\mathbb R}^n \times {\mathbb R}^n\rightarrow {\mathbb R}^{n^2}$.  By this notation we will state for convenience identity operators, $P_1(B)=B$ and $p_1(\bw)=\bw$.  Higher order operators follow similarly, but we will not need these here.

With this notation, we can proceed comparably to Eqs.~(\ref{iterate})-(\ref{fin}), by tracking iterations of the RC but with quadratic read-out, and with the terms in the $\circ$ product for use in building $\bW^{out}$ to be used in the read-out.  Let $\br_1=0$.   Then,
\begin{eqnarray}
\br_2 \circ \br_2 &=& (\bW^{in} \bx_1)\circ (\bW^{in} \bx_1) \nonumber \\
&=& P_2(\bW^{in},\bW^{in})p_2(\bx_1), \nonumber \\
\br_3\circ \br_3 &=& (A \bW^{in} \bx_1+ \bW^{in}\bx_2)\circ (A \bW^{in} \bx_1+ \bW^{in}\bx_2) \nonumber \\
&=&(A \bW^{in} \bx_1)\circ(A \bW^{in} \bx_1) +(A \bW^{in} \bx_1) \circ ( \bW^{in} \bx_2) +( \bW^{in} \bx_2)\circ (A \bW^{in} \bx_1)+( \bW^{in} \bx_2)\circ ( \bW^{in} \bx_2) \nonumber  \\
&=&P_2(A \bW^{in}, A \bW^{in})p_2(\bx_1,\bx_1)+P_2(A \bW^{in}, \bW^{in})p_2(\bx_1,\bx_2)+ \nonumber \\
 & &+P_2( \bW^{in}, A \bW^{in})p_2(\bx_2,\bx_1)+P_2( \bW^{in},  \bW^{in})p_2(\bx_2,\bx_2)\nonumber \\
&\vdots& \nonumber \\
\br_{k+1}\circ \br_{k+1}&=&\sum_{i=1}^k (A^{i-1} \bW^{in} \bx_{k+1-i})\circ (\sum_{j=1}^k A^{j-1} \bW^{in} \bx_{k+1-j}) \nonumber \\
&=& \sum_{i,j=1}^k P_2( A^{i-1} \bW^{in}, A^{j-1} \bW^{in}) p_2(\bx_{k+1-i},\bx_{k+1-j})  \\
&:=&{\mathbb A}_2 [{\mathbb X}_2]_k.
\end{eqnarray}

That is, ${\mathbb A}$ defined in Eq.~(\ref{Akry}) is $d_r\times k d_x$, and analogously, 
\begin{eqnarray}\label{A2}
{\mathbb A}_2=[&P_2(\bW^{in},\bW^{in}) & | P_2(A\bW^{in},\bW^{in}) | P_2(A^2\bW^{in},\bW^{in}) |... \nonumber \\
&...& |P_2(A^{k-1}\bW^{in},\bW^{in}) |P_2(\bW^{in},A\bW^{in}) |P_2(A\bW^{in},A\bW^{in}) |P_2(A^2\bW^{in},A\bW^{in}) |... \nonumber \\
&...&|P_2(A^{k-2}\bW^{in},A^{k-1}\bW^{in})| P_2(A^{k-1}\bW^{in},A^{k-1}\bW^{in}) ],
\end{eqnarray}
is a $d_r\times k d_x^2$ matrix.

Similarly, where ${\mathbb X}$ is a $(kdx)\times (N-k)$ matrix of data,  ${\mathbb X}_2$ is a $(kdx^2)\times (N-k)$ matrix of data, but quadratic forms, analogous to the $(kdx)\times (N-k)$ array  ${\mathbb X}$ from Eqs.~(\ref{fit}), (\ref{meq}), and $[{\mathbb X}_2]_k$ is the $k^{th}$ column.
\begin{equation}\label{X2}
{\mathbb X}_1=\begin{bmatrix}
 |& | & \vdots &  |\\ 
\bx_k& \bx_{k+1} & \ldots & \bx_{N-1}  \\ 
|& | & \vdots &  |\\ 
 \bx_{k-1}& \bx_k & \ldots &  \bx_{N-2} \\ 
|& | &\vdots & |\\ 
\vdots & \vdots & \vdots & \vdots \\
|& | &\vdots & |\\ 
\bx_{1} & \bx_{2} & \ldots &  \bx_{N-k}\\
|& | & \vdots &  |\\ 
\end{bmatrix},
{\mathbb X}_2=\begin{bmatrix}
 |& | & \vdots &  |\\ 
p_2(\bx_{k},\bx_{k})& p_2(\bx_{k+1},\bx_{k+1}) & \ldots & p_2(\bx_{N-1},\bx_{N-1})  \\ 
|& | & \vdots &  |\\ 
p_2(\bx_{k-1},\bx_{k})& p_2(\bx_{k},\bx_{k+1}) & \ldots & p_2(\bx_{N-2},\bx_{N-1}) \\ 
|& | &\vdots & |\\ 
\vdots & \vdots & \vdots & \vdots \\
|& | &\vdots & |\\ 
p_2(\bx_{1},\bx_{k})& p_2(\bx_{2},\bx_{k+1}) & \ldots & p_2(\bx_{N-k-1},\bx_{N-1})  \\ 
|& | & \vdots &  |\\ 
p_2(\bx_{k},\bx_{k-1})& p_2(\bx_{k+1},\bx_{k}) & \ldots & p_2(\bx_{N-1},\bx_{N-2})  \\ 
|& | & \vdots &  |\\ 
p_2(\bx_{k-1},\bx_{k-1})& p_2(\bx_{k+1},\bx_{k-1}) & \ldots & p_2(\bx_{N-2},\bx_{N-2})  \\ 
|& | & \vdots &  |\\ 
\vdots & \vdots & \vdots & \vdots \\
|& | & \vdots &  |\\ 
p_2(\bx_{1},\bx_{1}) & p_2(\bx_{2},\bx_{2}) & \ldots &  p_2(\bx_{N-k},\bx_{N-k})\\
|& | & \vdots &  |\\ 
\end{bmatrix},
\end{equation}
Now we write, ${\mathbb X}=\begin{bmatrix}{\mathbb X}_1 \\ {\mathbb X}_2\end{bmatrix}$.

With Eq.~(\ref{RR}) and Eq.~(\ref{newR}) in mind, and with $[\bR]_2$ the $2^{nd}$ column of $\bR$, being $[\bR]_2=\begin{bmatrix}\br_{k+1} \\ \br_{k+1}\circ \br_{k+1}\end{bmatrix}$, we generalize the VAR stated in Eq.~(\ref{arnoldi1} ).  The quadratic NVAR follows,

\begin{eqnarray}\label{arnoldiq} 
    \by_{\ell+1}&=& \bW^{out} [\bR]_2\nonumber \\
    &=& \bW^{out}_1  \sum_{j=1}^{\ell} \bA^{j-1} \bW^{in} \bx_{\ell-j+1}+ \bW^{out}_2\sum_{i,j=1}^\ell P_2( A^{i-1} \bW^{in}, A^{j-1} \bW^{in}) p_2(\bx_{\ell+1-i},\bx_{\ell+1-j}) \nonumber \\
    &=&\bW^{out}_1 \bA^{\ell-1} \bW^{in} \bx_1+\bW^{out}_1\bA^{\ell-2} \bW^{in} \bx_2 +\ldots + \bW_1^{out}\bA \bW_1^{in} \bx_{\ell-1} +\bW_1^{out} \bW^{in} \bx_{\ell} +\nonumber \\
 & &  +   \sum_{i,j=1}^\ell \bW_2^{out} P_2( A^{i-1} \bW^{in}, A^{j-1} \bW^{in}) p_2(\bx_{\ell+1-i},\bx_{\ell+1-j})    \nonumber \\
    &=& a_\ell\bx_1+a_{\ell-1}\bx_2 +\ldots + a_{2} \bx_{\ell-1} +a_{1}\bx_{\ell}+ \nonumber \\
    & & +    a_{2,(\ell,\ell)} p_2(\bx_1,\bx_1)+   a_{2,(\ell-1,\ell)} p_2(\bx_2,\bx_1)+   ...+ a_{2,(1,1)} p_2(\bx_\ell,\bx_\ell)        \label{NVAR1}   
\end{eqnarray}
with notation for the $\ell$ linear coefficient $d_x\times d_x$ matrices as before,
\begin{equation}\label{prod111}
    a_{j}=\bW^{out}_1 \bA^{j-1}\bW^{in}, \mbox{ } j=1, 2,...,\ell,
\end{equation}
and now we have $\ell^2$ quadratic term coefficient $d_x\times d_x^2$ matrices, 
\begin{equation}\label{prod2}
a_{2,(i,j)}=\bW^{out}_2 P_2(A^{i-1} \bW^{in}, A^{j-1} \bW^{in}), i,j=1...\ell.
\end{equation}
This is the generalization of the VAR equation coefficients written explicitly in Eq.~(\ref{prod1}) to these coefficient matrices of a quadratic NVAR that results a linear RC with Hadamard quadratic readout.
}

\bibliographystyle{plain}
\bibliography{bibit}
\end{document}